\documentclass[twocolumn,superscriptaddress,secnumarabic,amssymb, nobibnotes, aps, bm,pre,floatfix]{revtex4-1}
\usepackage[english]{babel}
\usepackage[latin9]{inputenc}
\usepackage{graphicx}
\usepackage{multirow}
\usepackage[usenames,dvipsnames]{xcolor}
\usepackage{bm}
\usepackage{mathtools}
\usepackage{amsmath}
\usepackage{rotating}
\definecolor{oceanboatblue}{rgb}{0.0, 0.47, 0.75}
\definecolor{orange}{rgb}{1,0.5,0}
\definecolor{goodgreen}{rgb}{0.1,0.5,0}
\definecolor{goodred}{rgb}{0.7,0,0}
\usepackage[colorlinks,urlcolor=goodgreen,citecolor=blue,linkcolor=goodred]{hyperref}
\usepackage{cleveref}
\usepackage{tikz}
\usepackage{tocvsec2}
\usepackage{scrextend}
\makeatletter
\newcommand{\ii}{\text{i}}
\newcommand{\ee}{\text{e}}
\renewcommand{\exp}[1]{\ee^{#1}}
\newcommand{\comment}[1]{}

\usepackage[normalem]{ulem}

\makeatother
\begin{document}
\title{Corner modes of the breathing kagome lattice: origin and robustness}
%%%%%%%%%%%
\author{M. A. J. Herrera}
\thanks{These authors contributed equally.}
\affiliation{Centro de F\'isica de Materiales (CFM-MPC) Centro Mixto CSIC-UPV/EHU,
20018 Donostia-San Sebasti\'an, Basque Country, Spain}
\affiliation{Donostia International Physics Center, 20018 Donostia-San Sebasti\'an, Spain}
%%%%%%%%%%%
\author{S. N. Kempkes}
\thanks{These authors contributed equally.}
\affiliation{Institute for Theoretical Physics, 3584 CS Utrecht, Utrecht University, Netherlands}
%%%%%%%%%%%
\author{M. Blanco de Paz}
\thanks{These authors contributed equally.}
\affiliation{Donostia International Physics Center, 20018 Donostia-San Sebasti\'an, Spain}
%%%%%%%
\author{A. Garc\'{i}a-Etxarri}
\affiliation{Donostia International Physics Center, 20018 Donostia-San Sebasti\'an, Spain}
\affiliation{IKERBASQUE, Basque Foundation for Science, Euskadi Plaza, 5, 48009 Bilbao, Spain}
%%%%%%%%%%%
\author{I. Swart}
\affiliation{Debye Institute for Nanomaterials Science, Utrecht University, 3584 CC Utrecht, Netherlands}
%%%%%%%%%%%
\author{C. Morais Smith}
\email{Correspondence to: Dario.Bercioux@dipc.org, C.demoraissmith@uu.nl}
\affiliation{Institute for Theoretical Physics, 3584 CS Utrecht, Utrecht University, Netherlands}
%%%%%%%%%%%
\author{D. Bercioux}
\email{Correspondence to: Dario.Bercioux@dipc.org, C.demoraissmith@uu.nl}
\affiliation{Donostia International Physics Center, 20018 Donostia-San Sebasti\'an, Spain}
\affiliation{IKERBASQUE, Basque Foundation for Science, Euskadi Plaza, 5, 48009 Bilbao, Spain}
\begin{abstract} 
We study the non-trivial phase of the two-dimensional breathing kagome lattice, displaying both edge and corner modes. The corner localized modes of a two-dimensional flake were initially identified as a signature of a higher-order topological phase but later shown to be trivial for perturbations that were thought to protect them. Using various theoretical and simulation techniques, we confirm that it does not display higher-order topology: the corner modes are of trivial nature. Nevertheless, they might be protected. First, we show a set of perturbations within a tight-binding model that can move the corner modes away from zero energy, also repeat some perturbations that were used to show that the modes are trivial. In addition, we analyze the protection of the corner modes in more detail and find that only perturbations respecting the sublattice or generalized chiral and crystalline symmetries, and the lattice connectivity, pin the corner modes to zero energy robustly. A destructive interference model corroborates the results. Finally, we analyze a muffin-tin model for the bulk breathing kagome lattice. Using topological and symmetry markers, such as Wilson loops and Topological Quantum Chemistry, we identify the two breathing phases as adiabatically disconnected different obstructed atomic limits.
\end{abstract}
\date{\today}
\maketitle
Among the two-dimensional lattice structures, the kagome lattice has gained increasing interest due to the variety of phases that it can describe~\cite{mekata2003kagome}. These phases range from frustrated antiferromagnets to spin liquids~\cite{spinliquid1,spinliquid2,spinliquid3,spinliquid4}. The kagome lattice has been realized in several experimental set ups, such as optical lattices~\cite{opticalkagome1,opticalkagome2,opticalkagome3}, mechanical, electrical and acoustic metamaterials~\cite{EzawaPRB,Ni2018,xue2018}, and even colloidal crystals~\cite{colloidalkagome1,colloidalkagome2}. Another interesting platform where the kagome lattice has also taken an important role is photonic crystals~\cite{li2019,Chen_2020,Kirsch_2021,li2019,Shen2021}. Concepts like band topology or bulk-boundary correspondence have allowed light propagation without backscattering even with disorder in bosonic systems~\cite{Joannopoulos2008reflectionfree,Shalaev2019backscatteringfree}. If the system has robust corner modes, they will behave as stationary cavity modes in the corners of the photonic crystal. The robustness against perturbations was claimed be due to higher-order topological protection~\cite{EzawaPRL,Kempkes2019,Proctor_2020,Ni2018}.

A recent experimental realization of the kagome lattice (and the one that inspires this work) is in the framework of artificially designed electronic lattices~\cite{Kempkes2019}. This technique has its origin in the manipulation of adatoms on metallic surfaces~\cite{hla2003single}. The idea is to confine the surface state of the metal, which behaves as a two-dimensional electron gas (2DEG), by means of a user-defined potential that patterns an anti-lattice. The theoretical framework used in this work is known as the muffin-tin technique.
In this article, we study the breathing kagome model~\cite{EzawaPRL}, which is a kagome lattice with alternating intra- and inter-cell hopping amplitudes. A key feature of this system is the realization of a phase that exhibits corner localized zero-energy states in finite-size samples. Throughout the paper, we will call this phase non-trivial since it does not correspond to the same setup as a trivial phase. However, it does not contain topologically protected features. We perform this choice to distinguish this special phase from the trivial one, which is a regular insulator that does not display any corner or edge modes~\cite{TQC}. This non-triviality will be addressed further in the text and will be given a proper definition by means of topological markers.

The protection of such corner modes was addressed in Ref.~\cite{Ni2018}. There, it was claimed that a tripartite generalization of the chiral symmetry protects the zero-energy corner modes in the breathing kagome model. Subsequently, this concept was adopted by Refs.~\cite{Kempkes2019, li2019realizing, Weiner2020, li2019, Li2020, Yang2020,Proctor_2021}. However, the authors of Ref.~\cite{vanMiert_2020} explicitly showed that the corner modes in the breathing kagome lattice can be moved away from zero energy by applying a local perturbation respecting the three-fold rotational symmetry, the mirror symmetries, and the generalized chiral symmetry. In other words: the generalized chiral symmetry alone does not protect the corner modes. In addition, it was shown that the corner modes of the breathing kagome lattice can be understood by a destructive interference solution, based on the concept of destructive interference presented in Refs.~\cite{Kunst_2017, Kunst_2018, Kunst_2019}.

Our main result here is that the two breathing phases correspond to different atomic limits, one with trivial polarization, and the other with non-trivial polarization. In the literature, we frequently find that the topological characterization using the electric polarization leads to classifying the system as a higher-order topological insulator (HOTI). However, we do not find any topological feature in the latter phase after applying other more specific tools. Hence, the electric polarization alone cannot be used as a topological marker. We characterized the two phases using  well-established and complementary
tools found in the literature, specifically the Wilson loop spectrum~\cite{alexandradinata2014wilson} and Topological Quantum Chemistry~\cite{TQC}. The same tools have been used as well in the field of topological photonic crystals~\cite{BlancodePaz_2019}, and acoustic metamaterials~\cite{Peri_2020}, confirming the universality of this techniques to diagnose topological phases.

This article is structured as follows: in Sec.~\ref{sec:1}, we introduce the simplest tight-binding formalism describing the breathing kagome lattice. We recall several concepts of group theory allowing the characterization of the symmetry properties of the breathing and non-breathing phase. In Sec.~\ref{sec:2}, we review the concept of generalized chiral symmetry according to the literature. We study its properties by introducing perturbation terms to the tight-binding Hamiltonian that respect/break both generalized chiral symmetry and spatial symmetries in order to determine whether the modes are (i) corner localized and (ii) pinned to zero energy. We find a set of rules that ensure the degeneracy and localization of the corner modes. In Sec.~\ref{sec:3}, we complement the previous section analyzing the corner modes in terms  destructive interference. Within this approach, we will confirm the results obtained in Sec.~\ref{sec:2}. Finally, in Sec.~\ref{sec:4} we investigate the breathing kagome lattice using a muffin-tin formulation, which accounts for all the possible hopping terms. The two phases of the system are characterized by computing the Wilson loop spectrum and by applying the topological quantum chemistry framework.

\section{The breathing kagome model\label{sec:1}}
%
%
%%%%%%%%%%%%
\begin{figure*}
\centering
\includegraphics[width=\linewidth]{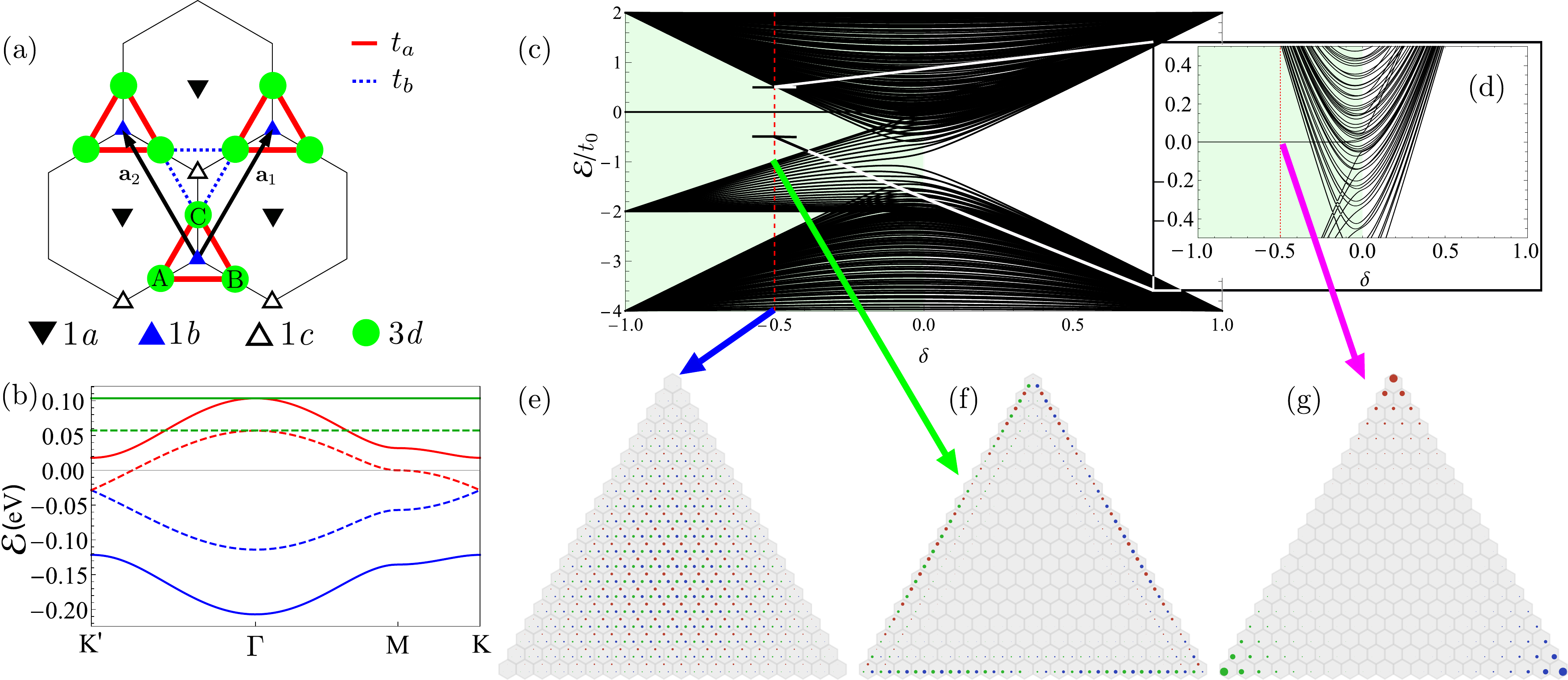}
\caption{\label{figure_one} The breathing kagome model. (a) Lattice model with three sublattice sites A, B and C, and hopping parameters $t_a$ and $t_b$. The black triangles represent the $1a$ Wyckoff position, the blue triangles represent the $1b$ Wyckoff position and the empty triangles represent the $1c$ Wyckoff position. The lattice sites fall on the $3d$ Wyckoff positions, represented with green circles. (b) Band structure of the kagome model for the breathing (solid lines) and non-breathing (dashed lines) phases. Notice that the non-breathing phase is gapless at the $\mathbf{K}$ and $\mathbf{K'}$ points. The corresponding values of $t_a=0.38t_b$ and $t_b=0.075\,\text{eV}$ are obtained from Ref.~\cite{Kempkes2019} by fitting the bands calculated within the tight-binding approach with the ones derived using the muffin-tin method. This corresponds to $\delta\approx-0.45$ and $t_0\approx52$~meV. (c) Spectrum of a finite-size lattice and (d) zoom-in for low-energy scale. For $\delta<0$, three-fold degenerated energy eigenvalues can be found pinned at zero. (e), (f), (g) Exponential decay of the wave function for $\delta=-0.5$ in the bulk, edge, and corner, respectively. In the last three panels, the size of the dots is set for convenience proportional to $|\psi|^{0.2}$.}
\end{figure*}
%%%%%%%%%%%
%
%

The breathing kagome is a two-dimensional lattice with alternating strong and weak hopping in a kagome pattern~\cite{EzawaPRL}. In Fig.~\ref{figure_one}(a), we show the unit cell and the choice of lattice vectors that we will consistently use in this work. The unit cell contains three sites, labelled A, B and C, respectively. When considering two-dimensional systems and within the single-particle description, the physics of both electrons and photons is identical. In this way, tight-binding methods can be applied to photons in the same way as they can be applied to electrons, regardless of the bosonic/fermionic nature of the particles. In the simplest tight-binding formulation, where we consider only nearest-neighbor hopping terms, the Hamiltonian in reciprocal space reads
\begin{align}\label{eq_Hamiltonian1}
\hat{\mathcal{H}}=\begin{pmatrix}
		0 & \!t_a\!+t_b\exp{\ii \mathbf{k}\cdot\mathbf{a}_3} & \!t_a\!+t_b\exp{\ii \mathbf{k}\cdot\mathbf{a}_2}\\
		t_a\!+t_b\exp{-\ii \mathbf{k}\cdot\mathbf{a}_3}&0& \!t_a\!+t_b\exp{\ii \mathbf{k}\cdot\mathbf{a}_1}\\
		t_a\!+t_b\exp{-\ii \mathbf{k}\cdot\mathbf{a}_2} & \!t_a\!+t_b\exp{-\ii \mathbf{k}\cdot\mathbf{a}_1} & 0
	\end{pmatrix},
\end{align}
with $t_a$ and $t_b$ the hopping parameters, $\bf k$ the crystal momentum, and ${\bf a}_{1,2}=(\pm\frac{1}{2}, \frac{\sqrt{3}}{2})$, and ${\bf a}_3={\bf a}_2-{\bf a}_1=(-1,0)$ the lattice vectors. The Hamiltonian~\eqref{eq_Hamiltonian1} is expressed in the basis $\Psi=\{\psi_\text{A},\psi_\text{B},\psi_\text{C}\}^\text{T}$, where $\text{T}$ represents the transposition. We show in Fig.~\ref{figure_one}(b) the band structure for a periodic lattice both in the breathing phase $(t_a\neq t_b)$ (solid lines) and in the canonical phase $(t_a=t_b)$ (dashed lines). Figure~\ref{figure_one}(c) presents the energy bands of a finite-size lattice obtained using the following  parametrization of the hopping amplitudes: $t_{a,b}=(1\pm\delta)t_0$ with $t_0<0$. The parameter $\delta$ is the breathing factor: it allows us to study the different phases of the breathing kagome lattice by changing its sign. Additionally, we consider a finite-size triangular flake of the breathing kagome lattice with a size of 630 lattice sites, or 20 unit cells along the side of the triangle. For $\delta=-1$, the fully dimerized case is recovered, and  the corner modes remain at zero energy.  Upon increasing the dimerization parameter they eventually hybridize with the bulk modes. Figure~\ref{figure_one}(d) reveals that the corner modes are truly pinned to zero energy, with no features in the spectrum. Figures~\ref{figure_one}(e)--(g) show the spatial localization of selected states for $\delta = -0.5$, revealing the existence of bulk, edge and corner modes, respectively. For the case of the corner modes | Fig.~\ref{figure_one}(g) | we notice that the wave function has non-zero weight in only one sublattice (A,B or C). The breathing kagome has two different phases, one featuring zero-energy corner-localized modes and edge modes ($\delta<0$) and one where such modes are absent ($\delta>0$), separated by a gapless one ($\delta=0$). 

In the following, we will give a short summary of the symmetry properties of the model. The canonical kagome lattice ($\delta=0$) belongs to the space group $p6mm$ (\#183 in the ITA~\cite{ITA}), characterized by a six-fold symmetry (point group $C_{6v}$) that closes the gap at the $\mathbf{K}$ and $\mathbf{K}'$ points in the first Brillouin zone | see dashed lines in  Fig.~\ref{figure_one}(b). After introducing the breathing distortion, the $C_6$ symmetry is broken, and we arrive at a different space group for the breathing kagome lattice, i.e., space group $p3m1$ (\#156 in the ITA~\cite{ITA}). This space group is a subgroup of $p6mm$ and has a three-fold rotation operation (point group $C_{3v}$), which opens the gap at the $\mathbf{K}$ and $\mathbf{K}'$ points | see solid lines in Fig.~\ref{figure_one}(b). The group/subgroup relation between $p6mm$ and $p3m1$ affects the naming of the Wyckoff positions~\cite{bcs1,bcs2,bcs3}. For this specific case, the $2b$ Wyckoff position of $p6mm$ splits into two non-equivalent Wyckoff positions of $p3m1$, i.e., $1b$ and $1c$, both with point group $C_{3v}$. This distinction will be crucial for studying the topological character of the bands, which will be discussed in Sec.~\ref{sec:4}, along with a better understanding of the closing and opening of a gap, in terms of group theory. Additionally, the symmetry of the $3c$ Wyckoff position, now called $3d$, reduces from $C_{2v}$ to $C_m$. Figure~\ref{figure_one}(a) shows these Wyckoff positions distributed in space. We have used symbols with the same symmetry as the point group of the Wyckoff position. In the case of the $3d$ Wyckoff position, we have used a circle for simplicity due to the reduced symmetry of this Wyckoff position.
%
%
%%%%%%%%%%%
\begin{figure*}
\centering
\includegraphics[width=.95\linewidth]{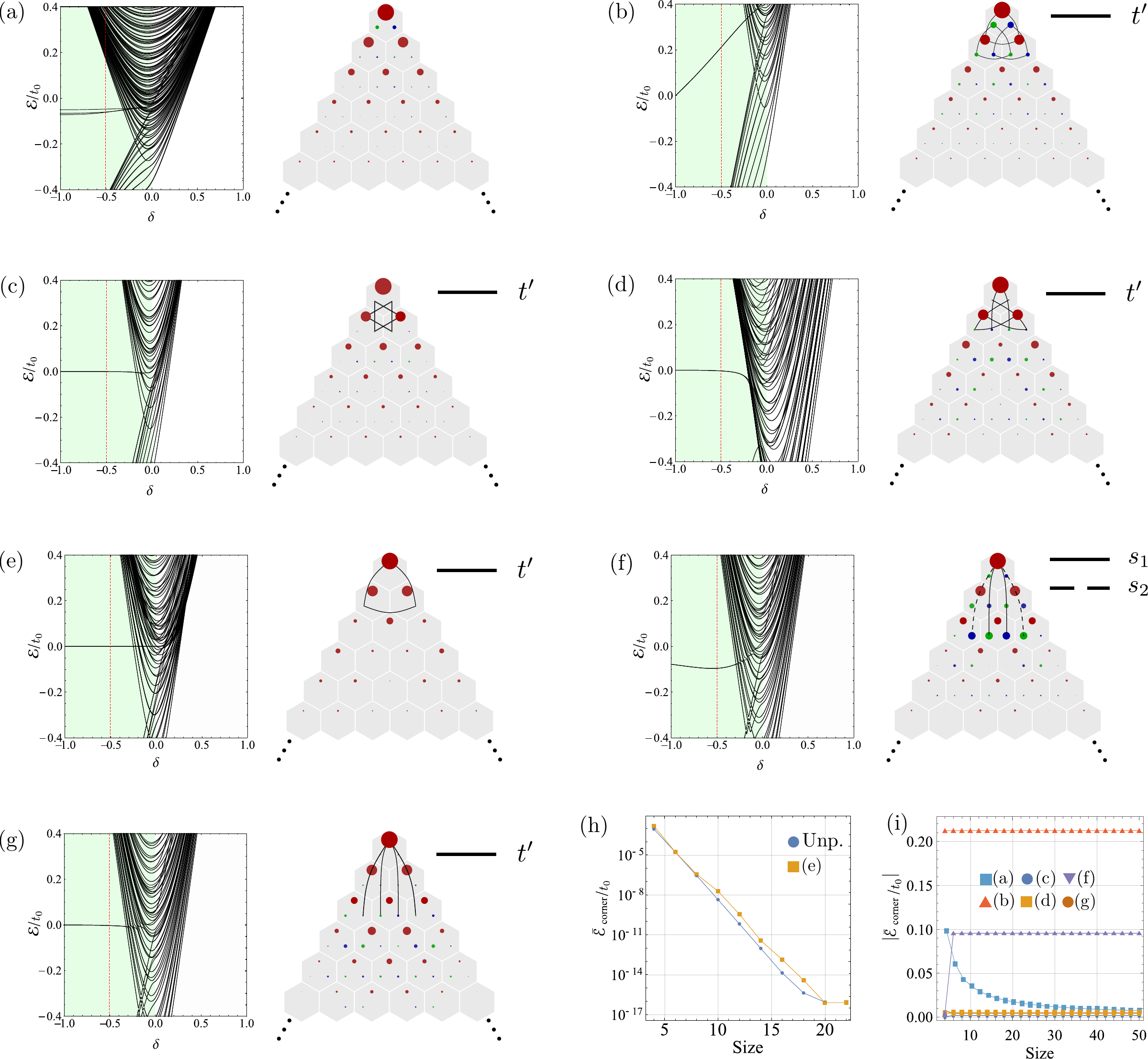}
\caption{\label{figure_two}
Set of perturbations that have been studied to detect the protection mechanisms of the corner modes. Each panel display the close-up of the spectrum plus the localization in real space of the wave function, thus, a visualization of the local density of states. (a) Random disorder in all lattice sites; (b) perturbation breaking generalized chiral symmetry, thus connecting lattice sites of the same species; (c) perturbation connecting second-order nearest neighbors; (d) perturbation connecting third-order nearest neighbors; (e) perturbation connecting fourth-order nearest neighbors; (f) long-range local perturbation with different sign; (g) long-range local perturbation with same sign; (h) and (i) influence of the size on the average energy, in absolute value, of the corner modes on each of the perturbations. Panel (h) has a logarithmic scale due to exponential localization.
}
\end{figure*}
%%%%%%%%%%%%
%
%
\section{The generalized chiral symmetry and its breaking\label{sec:2}}

The idea of the generalized chiral symmetry for the breathing kagome model follows the same line of reasoning as the chiral symmetry in the Su-Schrieffer-Heeger (SSH) model~\cite{SSH1979,Heeger,Asbth_2016}. It should be noted that chiral symmetry, also known as sublattice symmetry, is not a symmetry operation: instead of commuting with the Hamiltonian, chiral symmetry anti-commutes with it. Nevertheless, we will continue naming this operation chiral symmetry throughout the text, so as done in the literature. To introduce the generalized chiral symmetry for the case of the kagome lattice, we use the Bloch Hamiltonian~\eqref{eq_Hamiltonian1}, and define $\mathcal{H}_1=\hat{\mathcal{H}}$.
The kagome lattice is not a bipartite lattice; it has an odd number of lattice sites in the unit cell, as opposite to the one- and two-dimensional SSH models, which show an even number of lattice sites. 
Here, we repeat how to generalize the chiral symmetry for a unit cell containing three sites, in line with Ref.~\cite{Ni2018}. The generalized chiral symmetry is defined as some operator $\Gamma_3$ that satisfies
%
%
%%%%%%%%%%
\begin{subequations}\label{genchiralsym}
\begin{align}
\Gamma_3^{-1} \mathcal{H}_1 \Gamma_3& = \mathcal{H}_2,\label{genchiralsym_a} \\
\Gamma_3^{-1} \mathcal{H}_2 \Gamma_3 &=\mathcal{H}_3, \label{genchiralsym_b} \\
 \mathcal{H}_1+ \mathcal{H}_2+ \mathcal{H}_3&=0.\label{genchiralsym_c}
\end{align}
\end{subequations}
%%%%%%%%%%
%
%
When combining the last equation with the previous two, it follows that $\Gamma_3^{-1} \mathcal{H}_3 \Gamma_3 =\mathcal{H}_1$. Following this reasoning, the generalized chiral symmetry introduced in Eqs.~\eqref{genchiralsym} is completely analogous to the chiral symmetry of the SSH model~\cite{Asbth_2016}. However, in this case $[\mathcal{H}_1, \Gamma_3^3]=0$, which implies $\Gamma_3^3 = \mathbb{I}_3$ and the eigenvalues are given by $1, \text{exp}[\pm 2 \pi \text{i}/3]$. Therefore, up to a unitary transformation, we can write
%
%
%%%%%%%%%%
\begin{align}
\Gamma_3 = 
\begin{pmatrix}
 1 & 0 & 0 \\
 0 & \ee^{ 2\pi \text{i} /3} & 0 \\
 0 & 0& \ee^{- 2 \pi \text{i} /3}
\end{pmatrix}.
\end{align}
%%%%%%%%%%
%
%
Furthermore, we now have three eigenvalues to consider ($\mathcal{H}_1$, $\mathcal{H}_2$, and $\mathcal{H}_3$ each have the same eigenvalues $\epsilon_1$, $\epsilon_2$, and $\epsilon_3$, since the Hamiltonians differ by a unitary transformation). By taking the trace of Eq.~\eqref{genchiralsym_c}, we find
%
%
%%%%%%%%%%%%%%%
\begin{equation}\label{trace_GCS}
    \text{Tr}[\mathcal{H}_1+\mathcal{H}_2+\mathcal{H}_3]=3\text{Tr}[\mathcal{H}_1]=0, 
\end{equation}
where we used the first two lines of Eqs.~\eqref{genchiralsym} and the fact that the trace is cyclic. This means that the sum of the three eigenvalues vanishes, $\epsilon_1 + \epsilon_2 +\epsilon_3 =0$.

Now, the same reasoning could apply to the eigenstates. However, there is one crucial  difference. If $\mathcal{H}_1 | \psi \rangle = \epsilon_1 | \psi \rangle$, with $| \psi \rangle$ an eigenstate, the wave functions $\Gamma_3 | \psi \rangle$ and $\Gamma^2_3 | \psi \rangle$ are not necessarily also eigenstates of $\mathcal{H}_1$. In the SSH chain, this relationship is guaranteed by the relation $H_1= - H_2$. This does not hold for the generalized chiral symmetry, since  
%
%
%%%%%%%%%
\begin{equation}
    \mathcal{H}_1 \Gamma_3 |\psi \rangle = \Gamma_3 \mathcal{H}_2 |\psi \rangle,
\end{equation}
%%%%%%%%%
%
%
and since $| \psi \rangle$ is not per se an eigenstate of $\mathcal{H}_2$, it is not proven that $\Gamma_3 | \psi \rangle$ is an eigenstate of $\mathcal{H}_1$. Therefore, the generalized chiral symmetry does not work in the same way as the chiral symmetry, and there is no guarantee that a zero-energy mode will remain pinned to zero.

In the last part of this section, we will study the influence of perturbations on the electronic structure of a finite-size triangular flake, similar to what was done in Ref.~\cite{vanMiert_2020}. Note that this shape preserves the $C_{3v}$ symmetry. Figure~\ref{figure_two} sums up all the perturbations that we have studied, showing a close up of the spectrum around zero energy and the local density of states of the corner mode in the upper corner of the flake. To facilitate the visualization of the localization of the wave function at the corner, the size of the dots is proportional to $|\psi|^{0.2}$. We classify all these perturbations into four groups:
\begin{enumerate}
\item[(i)] global perturbation that breaks all possible spatial/crystalline symmetries, Fig.~\ref{figure_two}(a);
    \item[(ii)] global perturbation that breaks generalized chiral symmetry, Fig.~\ref{figure_two}(b);
     \item[(iii)] global perturbations that respect generalized chiral symmetry and the crystalline symmetries, and couple beyond nearest-neighbor sites, Figs.~\ref{figure_two}(c) to~\ref{figure_two}(e);
     \item[(iv)] local perturbations applied on the corners as in Ref.~\cite{vanMiert_2020}, Figs.~\ref{figure_two}(f) and~\ref{figure_two}(g).
\end{enumerate}

In Fig.~\ref{figure_two}(a), we add random on-site energies, ranging between 0 and $0.2t_0$, in all the lattice sites (bulk, edges and corners). We only show a single possible realization of random on-site energies, after finding similar results for several different disorder realizations. This perturbation breaks all possible spatial symmetries, while preserving the connectivity of the kagome lattice. This means that the generalized chiral symmetry is preserved in terms of connectivity, but spatial symmetries are no longer mapping the lattice to itself. We observe that the corner modes are neither pinned to zero energy nor degenerate; each one departs from zero at a different energy, even in the fully dimerized case. If we look at the localization of the mode around the corner, we see that the wave function has non-zero weight in the three sublattices, and that each circle has a different diameter as a consequence of the breaking of the symmetries. This may not be distinguished easily in the plot, but was confirmed numerically.

Figures ~\ref{figure_two}(b) to~\ref{figure_two}(g) show other types of perturbations: we introduce new hopping terms that change the connectivity of the lattice, while preserving both $C_{3v}$ and/or generalized chiral symmetry. The intensity of those hopping term has been set to the same value of the maximum random on-site energy ($0.2t_0$) used in Fig.~\ref{figure_two}(a). In Fig.~\ref{figure_two}(b), we show a perturbation that couples sites of the same sublattices, thus breaking generalized chiral symmetry. As soon as we depart from the fully dimerized case, the corner modes are no longer pinned at zero energy, but they are still degenerate, since the flake is $C_{3v}$ symmetric. Again, the wave function has nonzero weight in all the three sublattices, but this time the size of the circles is related by the mirror symmetry that maps the corner to itself (vertical mirror). Figures~\ref{figure_two}(c) to~\ref{figure_two}(e) show different choices of long-range hopping terms in increasing order of neighbor coupling (2nd, 3rd and 4th, respectively), which preserve both generalized chiral symmetry and $C_{3v}$ symmetries. In Fig.~\ref{figure_two}(c), the lattice site placed in the corner is unperturbed by this choice of hopping, and thus the spectrum is very similar to Fig.~\ref{figure_one}(d). However, the wave function shows nonzero weight in the three sublattices due to the different connectivity. Only the case in Fig.~\ref{figure_two}(e) respects the connectivity of the lattice sites with the same strength of hopping (closed triangle connecting A, B and C)~\footnote{We have further considered longer range hopping terms of this type. While these terms affect the localization energy of the corner modes, their exponentially localized nature as a function of the flake size is maintained, independently of the order of the hopping term. Thus, we recover the same behavior as in Fig.~\ref{figure_two}(i) but with a different slope.}. This connectivity is the same as in the unperturbed kagome lattice, but in a longer range. Indeed, the wave function shows nonzero weight in just one sublattice (to which the corner site belongs). The corner modes are tightly pinned to zero and are three-fold degenerate. Including longer range perturbations of this type gives rise to the same behavior. This is the same rule of localization that we described in Sec.~\ref{sec:1} for the unperturbed lattice. This perturbation leaves the corner modes untouched (for a sufficiently large sample).

In the case of Fig.~\ref{figure_two}(d), the perturbation also preserves the spatial symmetries, but couples the corner to the bulk and does not respect the connectivity of the kagome lattice. Hence, the wave function shows non-zero weight in different sublattices and the corner modes move away from zero energy (although not as quickly as for some of the other perturbations). 
The cases in Figs.~\ref{figure_two}(f) and~\ref{figure_two}(g) correspond to the perturbations introduced in Ref.~\cite{vanMiert_2020}, which are used as immunity checks for the robustness of the corner modes. These are long-range hopping amplitudes $s_1$ and $s_2$ applied locally at the corners. We have studied two different configurations: in Fig.~\ref{figure_two}(f) the spectrum is generated using $s_1=-s_2$, with $|s_1|=|s_2|=0.2t_0$, while the spectrum in Fig.~\ref{figure_two}(g) is generated using the same sign for the perturbations. The perturbation shown in Fig.~\ref{figure_two}(f) leads to degenerate modes, which are however no longer pinned to zero energy (not even in the fully dimerized case). When using the same sign for the perturbation [Fig.~\ref{figure_two}(g)], the modes are degenerate and located at zero energy in the fully dimerized case, but move away as $\delta$ increases.  In both cases, the wave function is delocalized over all three sublattices. These perturbations respect the generalized chiral symmetry and the $C_{3v}$ symmetry of the flake, but are applied only locally to the corners.

Finally, we present in Fig.~\ref{figure_two}(h) and Fig.~\ref{figure_two}(i) the evolution of the energy of the corner modes of the different perturbations with increasing size of the flake | the absolute value in the case of panel (i). We plot the average of the three closest-to-zero eigenvalues versus the number of unit cells along the edge of the flake. In certain cases, some of those values were negative or positive, sometimes some were degenerate, but in any case very close to zero. In Fig.~\ref{figure_two}(h), we show the influence of size on the unperturbed breathing kagome lattice and perturbation shown Fig.~\ref{figure_two}(e). These two systems are the only two with corner modes that fulfill the localization rule; hence, they exhibit a similar behavior. The differences between those two lines can be ascribed to different overlaps of the corner modes. This change of the overlap is associated with longer-range perturbations included in the model.

Figure~\ref{figure_two}(i) represents the remaining perturbations (the rest of the hopping terms and random on-site energies). Some of the perturbations pin the corner modes to values different from zero: perturbations (b) and (f) clearly pin the modes away from zero; perturbation (c) pins the modes very close to zero, while (d) and (g) are a bit away from zero. Finally, the light blue curve | perturbation (a) | shows the evolution of the energies of the flake with random on-site energies. In order to perform this calculation, we generated 20 configurations for each size, and then, we took the average of the energies of the corner modes. We see a $1/N$ evolution of the eigenvalues, where $N$ is the total number of unit cells. This precise evolution suggests that for small sizes, the corner modes are not pinned to zero due to poor localization. In the thermodynamic  limit, where $N\rightarrow\infty$, the long tails of the modes will eventually remain isolated, even if the intercell hopping is not zero, and thus the corner modes reach the on-site energies of the corner.

In addition to these results, we refer to the supplementary material of Ref.~\cite{Kempkes2019} for a similar study of the perturbation of the corner modes. In the field of photonic crystals, we refer to Ref.~\cite{Proctor_2020} for a similar analysis of the robustness of corner modes in a photonic breathing honeycomb lattice. In addition, in the field of plasmonics, we refer to Ref.~\cite{Proctor_2021} for a realization of the breathing kagome lattice in such framework, as well as for a study of robustness of corner modes. We refer to Ref.~\cite{El_Hassan_2019} and the supplementary material of Ref.~\cite{kagomewaveguide} for a complementary study of the robustness of the corner modes in waveguide arrays. Finally, other geometries may also host robust corner modes, as it is the case of Ref.~\cite{han2020lasingC4,gong2021topologicalC4,ota2019photonic,chen2019direct}. In these geometries, chiral symmetry, in addition to spatial symmetries, yield to a further protection of the corner modes.

\section{Destructive interference interpretation of the corner modes\label{sec:3}}

A different  way to interpret the zero modes in the breathing kagome lattice is by considering them as due to destructive interference~\cite{Kunst_2017,Kunst_2018, Kunst_2019}. To illustrate this approach, we will  follow Ref.~\cite{Kunst_2017}: we start by considering  the case of a one-dimensional (1D) bipartite lattice with two sites in the unit cell, A and B, such as the SSH chain. When considering a Hamiltonian in which the A sites only couple to one B site, it is possible to find a wave function that completely localizes on the A sublattice due to destructive interference. 

In this perspective, the non-trivial phase of the SSH model can be understood in terms of destructive interference. To describe the zero-energy modes of the breathing kagome lattice within this approach, we will start by analyzing  destructive interference in a SSH-like model.
In fact, the model of destructive interference admits an analytical solution if the chain starts and ends with the same type of lattice site, \emph{e.g.} A~\cite{Kunst_2019,footnote}.

In order to find the zero-energy wave function that interferes destructively on the B sublattice, we use the \emph{ansatz}
%
%
%%%%%%%%%%%%%%
\begin{equation}\label{wf_di}
    |\psi \rangle = N_i \sum_m^M r_i^m c_{\text{A}_i,m}^\dagger |0\rangle,
\end{equation}
%%%%%%%%%%%%
%
%
where $r$ is a complex number describing the  wave function decay, $N_i$ a normalization constant, $M$ the total number of unit cells, and $c_{\text{A}_i,m}^\dagger$ creates an electron on an A-site of cell $i$.

\begin{figure}
\centering
\includegraphics[width=\columnwidth]{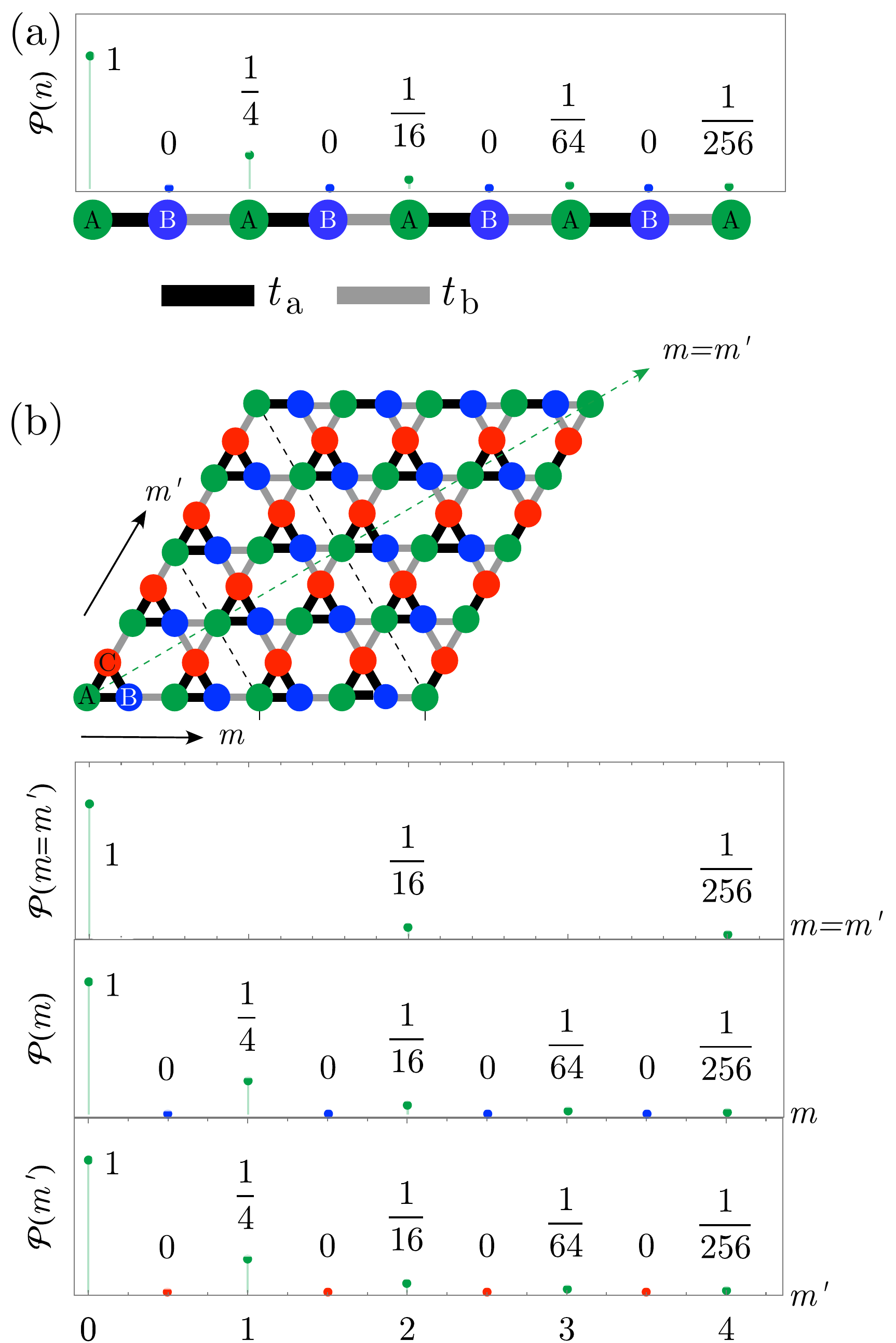}
\caption{\label{SSH_kagome} Destructive interference models. (a) In the case of the 1D chain, the probability is finite only in the A sublattice site and decays as $\left(-t_a/t_b\right)^{2n}$, where $n$ is the unit-cell index. (b) The breathing kagome rhombus. Due to destructive interference, there is a wave function that has zero amplitude on the B and C sublattices and a finite amplitude on the A, where the probability decays as $\left(-t_a/t_b\right)^{(2m')}$ along $m$ (and analogously along $m'$), whereas it decays as $\left(-t_a/t_b\right)^{2(m+m')}$ along $m+m'$. In all the panels, we have set $t_2=2t_1$.}
\end{figure}

If the A sites only couple to the B sites and vice versa, the Hamiltonian for this 1D lattice with open boundaries reads
%
%
%%%%%%%%%%%%%%%
\begin{equation}\label{ham_SSH}
  \bm{H}_\text{1D}=  \begin{pmatrix}

 e_\text{A} & t_\text{A,B} & 0 & 0&0\\
 t_\text{A,B}^\dagger & e_\text{B} & t_\text{B,A} &  0 &0  \\
 0 & t_\text{B,A}^\dagger  & e_\text{A} & \cdots & 0\\
 0 & 0 & \vdots & \ddots & t_\text{B,A} \\
 0 & 0 & 0 & t_\text{B,A}^\dagger & e_\text{A}
\end{pmatrix},
%\right),
\end{equation}
%%%%%%%%%%%%%%
%
%
where $e_\text{A,(B)}$ is the on-site energy for the A (B) lattice site and, $t_\text{A,B}$ and $t_\text{B,A}$ are the intra- and inter-hopping terms between the lattice sites. We can rewrite  Eq.~\eqref{wf_di} as
%
%
%%%%%%%%%%%%%%
\begin{equation}\label{wfdeint}
    |\psi\rangle = 
\begin{pmatrix}
 1, 0, r, 0, r^2, 0, r^3,
 \ldots
\end{pmatrix}^\text{T}, 
\end{equation}
%%%%%%%%%%%%%%
%
%
localized only on the  A sites, where we have omitted the normalization factor. The action of Hamiltonian~\eqref{ham_SSH} on this wave function is
%
%
%%%%%%%%%%%%%
\begin{align}
 &   \bm{H}_\text{1D}  |\psi\rangle  =  \\
 &
 \begin{pmatrix}
 e_\text{A}, t_\text{A,B}^\dagger + r t_\text{B,A},r e_\text{A},
 r (t_\text{A,B}^\dagger + r t_\text{B,A}),
 \hdots,
 e_\text{A} r^M 
\end{pmatrix}^\text{T} \nonumber
\!\!\!.
\end{align}
%%%%%%%%%%%%%
%
%
From this equation, it is clear that if $t_\text{A,B}^\dagger + r t_\text{B,A} = 0$, the wave function~\eqref{wfdeint} is an eigenstate of $\bm{H}_\text{1D}$ with eigenvalues $e_\text{A}$. It has the property that the weight on the B sites is $0$ and there is a decaying wave function with energy $e_\text{A}$ only on the A sites. We find $r= |- t_\text{A,B}^\dagger / t_\text{B,A}|$, and this mode is localized on the left of the chain if $r <1$, and on the right if $r>1$. In the case of  the SSH model, $e_\text{A}=e_\text{B}=0$, $t_\text{A,B}=t_a$ and $t_\text{B,A}=t_b$. We find $r=|- t_a/t_b|$, leading to the well-known localization of the zero mode on one side of the lattice~\cite{Kunst_2019}. This is true if the unbroken cell is on the right edge; the condition is reversed if the unbroken cell is on the opposite edge. A sketch of this wave function is given in Fig.~\ref{SSH_kagome}(a). This feature seems to indicate that once the lattice with open boundaries is ``long enough", these exact solutions of the wave function can be used to describe the zero modes of the SSH (even though in the SSH model the sites at the beginning and end of the chain are different). Note that these zero modes are now only present when $t_a<t_b$ (the non-trivial phase) because we can then map the zero mode of the SSH model to the one at the end of the chain discussed above. This cannot be done in the trivial phase, where the eigenstate is not starting at the end of the chain~\cite{Kunst_2019}. In this perspective, one does not need to invoke chiral symmetry and also when the on-site energy of a site is increased to $E=\epsilon$~\cite{Asbth_2016}, there will still be these exponentially decaying modes at energy $\epsilon$.

We now follow the analysis in terms of destructive interference  to the breathing kagome model~\cite{Kunst_2018}; the two-dimensional nature of the wave function leads to two indices $m$ and $m'$ in Eq.~\eqref{wf_di}. The wave function is therefore
%
%
%%%%%%%%%%%%%
\begin{equation}
    |\psi \rangle = N_i \sum_m^M \sum_{m'}^{M'} r_i^m  r_i^{'m'} c_{\text{A}_i,m,m'}^\dagger |0\rangle.
\end{equation}
%%%%%%%%%%%%%
%
%
The Hamiltonian for the breathing kagome lattice can be expressed as parallel 1D chains coupled to each other via the intermediate site C, Fig.~\ref{SSH_kagome}(b). The Hamiltonian for this breathing kagome rhombus reads 
%
%
%%%%%%%%%%%%%
\begin{equation}
     H^{M,M'}=  \begin{pmatrix}
 \bm{H}_\text{1D} & \bm{t}_\text{AB,C} & 0 & 0&0\\
 \bm{t}_\text{AB,C}^\dagger & \bm{e}_\text{C} & \bm{t}_\text{C,AB} &  0 &0  \\
 0 & \bm{t}_\text{C,AB}^\dagger  & \bm{H}_\text{1D} & \cdots & 0\\
 0 & 0 & \vdots & \ddots & \bm{t}_\text{C,AB} \\
 0 & 0 & 0 & \bm{t}_\text{C,AB}^\dagger & \bm{H}_\text{1D}
\end{pmatrix},
\end{equation}
%%%%%%%%%%%%%
%
%
where $\bm{H}_\text{1D}$ is the same as for Eq.~\eqref{ham_SSH} with $t_\text{A,B}=t_a$ and $t_\text{B,A}=t_b$, $\bm{e}_\text{C}$ is the matrix of the on-site energy of the site C, and $\bm{t}_\text{AB,C}$ and $\bm{t}_\text{C,AB}$ are the rectangular matrices containing the hopping elements connecting the 1D chains to the C sites. These are  given by
%
%%%%%%%%%%%%

%%%%%%%%%%%%
%
%
%
%
%%%%%%%%%%%%%%%%
\begin{subequations}
\begin{equation}
      \bm{t}_\text{AB,C}=  \begin{pmatrix}
t_a & 0 & 0& 0\\
 t_a & 0 & 0 &  0   \\
 0 & t_a  & 0 & 0 \\
 0 & t_a & \cdots &  0 \\
 0 & \vdots & \ddots & 0 \\
 0 & 0 & 0 & t_a
\end{pmatrix},
\end{equation}
%%%%%%%%%%%%
%
%
and
%
%
%%%%%%%%%%%%
\begin{equation}
\bm{t}_\text{C,AB}^\dagger=  
\begin{pmatrix}
 t_b & 0 & 0& 0\\
 0 & t_b &  &  0   \\
 0 & t_b  & 0 & 0 \\
  0 & \vdots & \ddots & 0 \\
 0 & 0 & 0 & t_b\\
 0& 0 & 0 &t_b
\end{pmatrix}.
\end{equation}
\end{subequations}
%%%%%%%%%%%%
%
%
In this way, the coupling between A and B or C is alternating $t_a$ and $t_b$. Using the same analysis as before, we observe that $t_a+ r t_b=0$ and $t_a+r' t_b=0$ for these exact wave functions, leading to
%
%
%%%%%%%%%%%%%%
\begin{equation}
    |\psi \rangle = N_i \sum_m^M \sum_{m'}^{M'} \left( \frac{-t_a}{t_b}\right)^m  \left(\frac{-t_a}{t_b}\right)^{m'} c_{\text{A}_i,m,m'}^\dagger |0\rangle.
\end{equation}
%%%%%%%%%%%%%
%
%
The real amplitude of such a wave function is shown in Fig.~\ref{SSH_kagome}(b). The rhombus-shaped flake, adopted from Ref.~\cite{xue2018}, allows having the same sublattice in each corner, in order to follow the same approach as in the SSH model explained previously. The three lower panels of  Fig.~\ref{SSH_kagome}(b) show the amplitude of the wave function along three different directions inside the flake. Along $m$ and $m'$ the weight of the wave function in the sublattice different from the one in the corner is always zero. The case of $m=m'$ is a consequence of the geometry, since we only find sublattice sites of the same kind as in the corner. Within this setup, we can generalize the hopping parameters connecting the sites by making them different. However, this will only add to the complexity of the model without changing the physics. The key point is that we can always find a solution for a decaying wave function with coefficients determined analytically.

%
%
%%%%%%%%%%%%%%%%
\begin{figure}[h!]
\centering
\includegraphics[width=1.0\columnwidth]{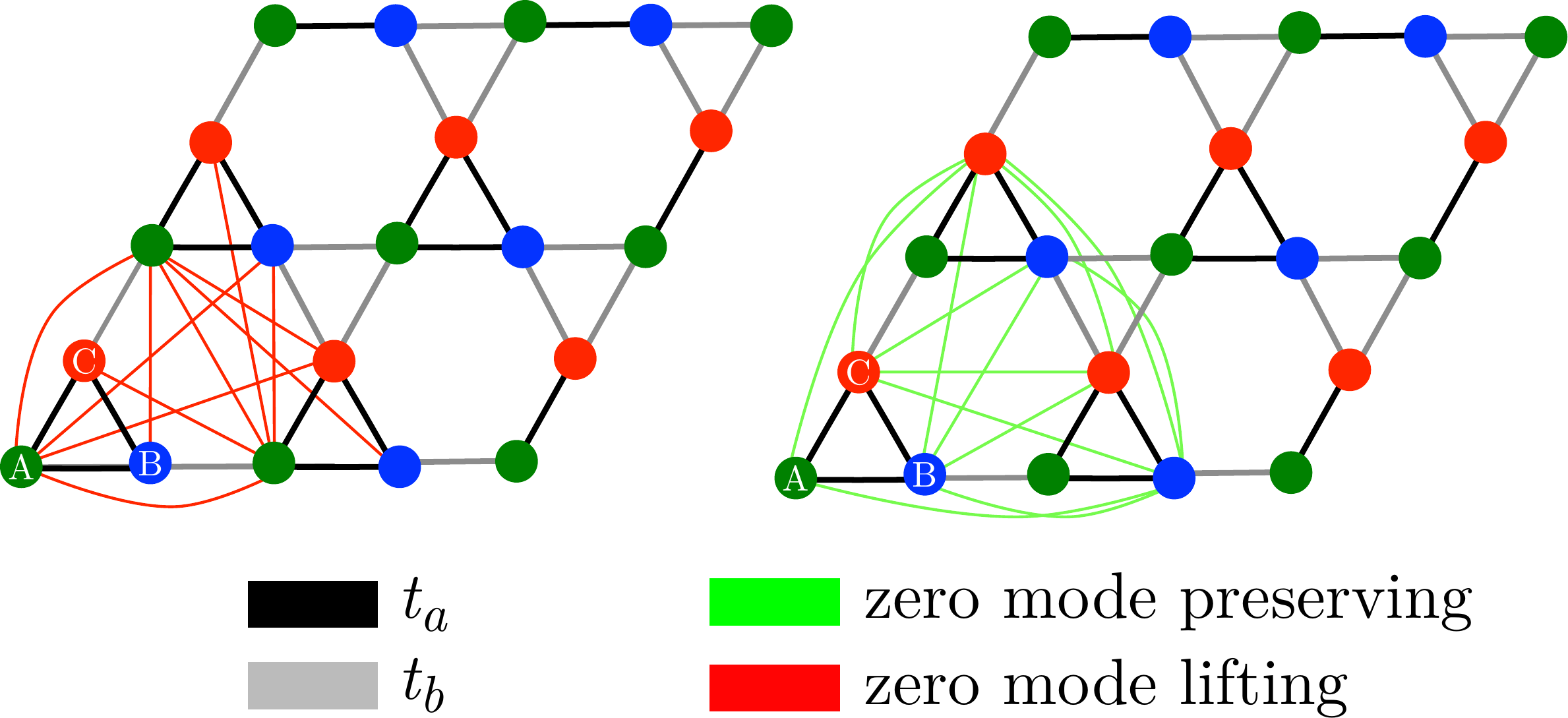}
\caption{\label{figure_kagome_rhombus} Set of hopping terms that lift (a) or preserve (b) the corner modes pinned to zero energy. Although they are only shown on the lower-left corner of the flake, these hopping terms can be established between any $m$ and $m'$ layers.}
\end{figure}
%%%%%%%%%%%%
%
%
A more interesting approach is to determine which additional hopping terms preserve the corner mode at zero energy. We consider all the hopping parameters indicated in Fig.~\ref{figure_kagome_rhombus} (see Appendix~\ref{appA} for the explicit expression of the Hamiltonian for a lattice containing 21 sites). For simplicity, we show in Fig.~\ref{figure_kagome_rhombus} only the hopping terms between the first three unit cells in the bottom left corner of the flake, but they extend to all the lattice. It turns out from the analysis on this lattice (in which we placed an A site in each of the corners of this breathing kagome rhombus~\cite{Kunst_2018}) that only the hopping terms indicated in green  preserve the energy of the corner modes, whereas the hopping terms in red change the energy of the corner modes (\emph{i.e.}, there is no consistent solution when we include the red hopping terms). To summarize, all hopping terms between the sites B and C preserve the corner mode energy and additionally one can connect A and B and A and C in the direction $m$ and $m'$, respectively (in the same way as in the SSH chain). However, one cannot connect an A site with another site (A, B or C) when these sites are in different chains: $m$ and $m'$ are both different. This analysis is fully consistent with the numerical analysis for a lattice containing 630 sites presented in Fig.~\ref{figure_two}. Note that in order to have zero modes in the triangle (with a different sublattice in each corner), we can only have the NNN hopping terms along $m$ (connecting A to B) or $m'$ (connecting A to C); all other perturbations will remove the zero mode since they connect the sublattice of the corner mode with a different site and hence the destructive interference is gone. We note in passing that the destructive interference method has been recently extended to the case of lattice systems characterized by a non-Hermitian Hamiltonian~\cite{Wong_2021}.
%
%
%%%%%%%%%%%%%%
\section{The muffin-tin method\label{sec:4}}
In this final section, we will analyse the experimentally realized breathing kagome lattice (Ref.~\cite{Kempkes2019}) in more detail. The experiment has been theoretically analyzed with two complementary theoretical approaches: the muffin-tin method and an extended tight-binding approach.  The former method describes a specific class of experiments, where a 2DEG on a surface of noble metals is patterned by molecules or atoms arranged in a precise and periodic fashion~\cite{khajetoorians2019creating,Kempkes2019,Freeney_2020,park2009making}. Specifically, in the experimental set up of Ref.~\cite{Kempkes2019}, the 2DEG is the surface state hosted by the (111) surface of Cu, and it was decorated  with a set of CO molecules adsorbed at certain positions, with the help of the tip of a scanning tunneling microscope \cite{Freeney_2020}. The muffin-tin method does not involve atomic orbitals or species, nor chemical bonds between them. The lattice sites are built with artificial interacting quantum dots (also known as artificial atoms~\cite{stilp2021artificial_atoms}) connected by hopping amplitudes which are always long-range and modeled by potential wells or barriers. This long-range hopping amplitudes can be fitted to nearest, next-nearest, etc, hopping terms in a tight-binding model. This property suggests that the muffin-tin method always takes into account all the possible hopping terms between all the lattice sites, namely those respecting generalized chiral symmetry and those which do not. Only the spatial symmetry of the potential will affect the properties of the 2DEG. 
\subsection{Muffin-tin potentials for canonical/breathing kagome lattices}
To study the breathing kagome lattice, we have considered three different configurations of CO molecules, accounting for the canonical gapless phase and the two breathing ones. Each molecule is modeled by a cylinder of radius $a$ and height $V_0$ placed at position $\mathbf{r}_n$:
%
%
%%%%%%%%%%%%%%%
\begin{align*}
V_n(\mathbf{r})=
 \begin{cases}
    V_0>0 & \text{if}~| \mathbf{r}-\mathbf{r}_n|<a,\\
    0 & \text{otherwise}.
 \end{cases}
\end{align*}
%%%%%%%%%%%%%
%
%
%
%
%%%%%%%%%%%%%%%
\begin{figure*}%[!h]
\centering
\includegraphics[width=\linewidth]{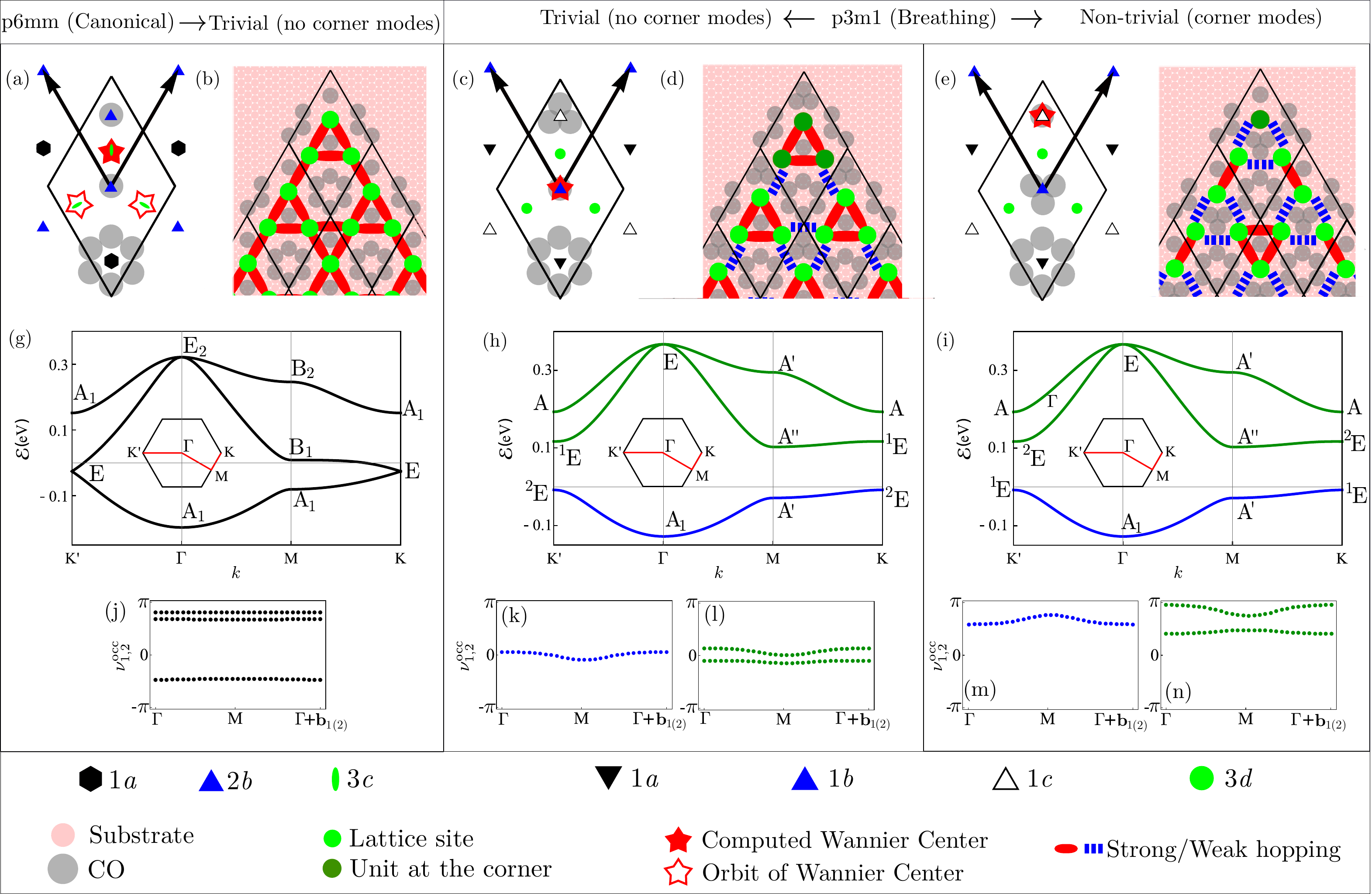}
\caption{Summary of the results obtained for the muffin-tin calculation of the canonical/breathing kagome lattice. The overall legend affects the three cases, while each column has its own legend for the Wyckoff positions, which are called differently due to the group/subgroup relation. Potential wells defining the unit cells for the non-breathing (a) and breathing phases (c), (e). The Wyckoff positions have been represented with an element with the same symmetries as in the point group associated with each Wyckoff position: hexagon, $C_{6v}$, triangle, $C_{3v}$, and ellipse, $C_{2v}$. In the case of the $3d$ Wyckoff position, we have used a circle for simplicity due to the reduced symmetry of this Wyckoff position (the point group associated is $C_m$, which includes just mirror and identity). Upper corner of a finite size sample of kagome lattice; non-breathing (b) and breathing phases (d), (f). Panels (g) to (i) show the corresponding band structures plus the irrep assignment at each high-symmetry point. Finally, panels (j) to (n) show the Wilson spectra obtained for the first three bands of each configuration.  \label{fig:UC_choice+bands}}
\end{figure*}
%%%%%%%%%%%%%%%
%
%
The full landscape is the superposition of the potential of each molecule. The design of the potential well is done by placing CO molecules forming the negative image of the lattice (a muffin-tin). Once the potential well defining the unit cell is built, the full lattice is constructed by translating it along the direct lattice vectors $\mathbf{a}_1$ and $\mathbf{a}_2$. We will be able to work with $s$-like or $p$-like orbitals, which allow us to study more complex interactions~\cite{slot2019p,Freeney_2020}. Practically, this is achieved by changing the size of the potential wells, which brings the energy levels up or down. In Figs.~\ref{fig:UC_choice+bands}(a),~\ref{fig:UC_choice+bands}(c) and~\ref{fig:UC_choice+bands}(e), we show the choice of unit cells that have been used to build the corresponding potential that reproduces the canonical kagome lattice, and the two breathing phases, respectively. We refer the reader to App.~\ref{App:instructions} for a step-by-step tutorial on how to reproduce the canonical and breathing phases using CO molecules on top of a Cu (111) surface. Many other choices can be realized, either by changing the molecule setup, or choosing different lattice vectors or different origins. We use configurations symmetric with respect to the mirror plane $m_{\bar{1}1}$ centered 
either at the $1b$ or at the lower $2b$ Wyckoff positions, depending on whether  we work with $p3m1$ or $p6mm$, respectively. Such geometric locus represents the center of mass of the three lattice sites inside the choice of unit cell, as well as the center of ``positive" charge~\footnote{This will be useful when we talk about bulk polarization in terms of relative displacement of the charge centers.}.

Once the potentials are built up, we solve the Schr\"odinger equation by expanding the potential in Fourier components in reciprocal space~\cite{Ashcroft76}. By obtaining the coefficients of such expansion, we can reconstruct the band structure and the Bloch wave functions for the three different configurations. In Figs.~\ref{fig:UC_choice+bands}(g), \ref{fig:UC_choice+bands}(h), and \ref{fig:UC_choice+bands}(i), we show the band structures along the high-symmetry path. In the bulk, the two setups of molecules are related by a $m_{11}$ mirror passing through the upper lattice site (geometric transformation). This explains why the eigenvalues of the two phases are the same. We nevertheless expect the eigenstates to behave differently, so we will distinguish these two phases via topological and symmetry markers, such as Wilson loops, bulk polarization, and Topological Quantum Chemistry.

Within the single particle picture, we may find a straightforward analogy with  photonic crystals, since the mathematical description of a muffin-tin experiment and the setup of a  two-dimensional photonic crystal has many things in common. On the one hand, with periodic boundary conditions, both muffin-tin potentials and  photonic crystals are solved by means of a plane-wave expansion of a differential secular equation: the Schr\"odinger equation in the former, with a periodic potential, and Maxwell's equation in the latter, with a periodic dielectric function~\cite{mpb,BlancodePaz_2019}. In both cases, we end up with an eigenvalue problem for the coefficients of such expansion. On the other hand, with open boundary conditions, such differential equations are solved inside a finite-size domain.
\subsection{Wilson spectrum analysis of the phases: Wannier center and bulk polarization}
%We study the topological properties of the breathing kagome lattice using the Wilson loop operator. This method is widely used in the literature to distinguish topological phases~\cite{BlancodePaz_2019,alexandradinata2014wilson,bercioux2018topological}. The spectrum of the Wilson operator allows to determine the topological character of a band structure, depending on its behavior. If the Wilson spectrum shows a winding as a function of momentum $k$, the system is topological, and the Wilson spectrum is connected to the value and sign of the corresponding Chern number characterizing the band structure. Conversely, if the Wilson spectrum is mapped to a constant value equal to zero, then the band structure shows trivial topology and the system is in a trivial atomic limit. If the Wilson spectrum can be mapped to any other constant value, the system is in an obstructed atomic limit. We will use these definitions later in the text (see sec. \ref{sec:4_TQC}). 

We study the topological properties of the breathing kagome lattice using the Wilson loop operator. This method is widely used in the literature to distinguish topological phases~\cite{BlancodePaz_2019,alexandradinata2014wilson,bercioux2018topological}. The spectrum of the Wilson operator allows to determine the topological character of a band structure, depending on its behavior. If the Wilson spectrum shows a winding as a function of momentum $k$, the system is topological, and the Wilson spectrum is connected to the value and sign of the corresponding Chern number characterizing the band structure. Conversely, if the Wilson spectrum is mapped to a constant value then the band structure shows trivial topology and the system is in an atomic limit. We will use these definitions later in the text (see sec. \ref{sec:4_TQC}). 

In order to introduce the Wilson spectrum, we first define the Wilson loop operator as the path ordered integral of the Berry connection along a certain path $\ell$. For an isolated band, the Wilson loop operator between $k$ points  $(k_1,k_2)$ and $(k_1+2\pi,k_2)$ is expressed as:
%
%
%%%%%%%%%%%%%
\begin{equation}
W^{n}_{(k_1+2\pi,k_2)\leftarrow (k_1,k_2)}=\mathcal{P}\text{ exp}\left\{-\text{i}\int_\ell d\ell\cdot\mathbf{A}_{n}\right\},\label{eq:WL}
\end{equation}
%%%%%%%%%%%
%
%
where $n$ is the band index, the symbol $\mathcal{P}$ represents path ordering operation and $\ell$ is the path between points $(k_1,k_2)$ and $(k_1+2\pi,k_2)$. The Berry connection is defined as 
%
%
%%%%%%%%%
\begin{align*}
\mathbf{A}_n(k_1,k_2)=-\ii\langle u_n(k_1,k_2)|\boldsymbol{\nabla}_{\mathbf{k}}|u_n(k_1,k_2)\rangle,
\end{align*}
%%%%%%%
%
%
where $|u_n(k_1,k_2)\rangle$ is the periodic part of the Bloch eigenfunction.

We work with the discrete version of Eq.~\eqref{eq:WL} by discretizing the reciprocal space along the two reciprocal space directions~\cite{BlancodePaz_2019}, with $N_k$ reciprocal lattice points along each direction. The Wilson line connecting two momenta along the reciprocal space vector $\mathbf{b}_1$ is $W^{n}_{(k_1+\delta k,k_2)\leftarrow (k_1,k_2)}=\langle u_n(k_1+\delta k,k_2)|u_n(k_1,k_2)\rangle $, so the total Wilson loop from $\Gamma$ to $\Gamma+\mathbf{b}_1$ is just the product of all these contributions,
%
%
%%%%%%%%%%%%%%
\begin{gather}
W^{n}_{(k_1+2\pi,k_2)\leftarrow (k_1,k_2)}=W^{n}_{\mathbf{b}_1}(k_2)=\nonumber\\
=\langle u_n(k_1+2\pi,k_2)|\prod_{j=1}^{N_k-1}\mathcal{P}(k^j_{1},k_2)|u_n(k_1,k_2)\rangle,\label{eq:def_WL}
\end{gather}
%%%%%%%%%%%
%
%\newline
where $k^j_{1}=j|\mathbf{b}_1|/N_k$ for $j=1,...,N_k-1$ and $\mathcal{P}(k_{1},k_2)$ is the projection operator $\mathcal{P}(k_1,k_2)=|u(k_1,k_2)\rangle\langle u(k_1,k_2)|$. If we are dealing with a composite group of $N_\text{occ}$ occupied bands, Eq.~\eqref{eq:def_WL} still applies, but the Wilson lines have band indices, so the Wilson line/loop becomes non-Abelian: $W^{mn}_{\bf k+\delta k\leftarrow k}=\langle u^m_{\bf k+\delta k}|u^n_{\bf k}\rangle$, where $m$ and $n$ range in all the occupied bands. At this point, the path ordering operation is crucial, since we are dealing with non-Abelian Wilson lines. Once the non-Abelian Wilson loop operator is built, we take the phases of the spectrum of this matrix to arrive at an equivalent result as given by Eq.~\eqref{eq:def_WL}.

Since the Wilson loop along $\mathbf{b}_1$ is a function of $k_2$, we can evaluate the Wilson loop for all the steps in the discretization along $k_2$. This is called tha Wilson spectrum, and it is related to the shifting of the Wannier center along the $\mathbf{a}_2$ direction. Due to $m_{\bar{1}1}$ symmetry, the Wilson spectrum along $\mathbf{b}_1$ is the same as along $\mathbf{b}_2$, and so will be the Wannier center~\cite{bercioux2018topological}. The position of the Wannier center is totally equivalent to the value of the bulk polarization, since the Wannier center represents the center of the negative electronic cloud. For $C_n$-symmetric insulators, the bulk polarization is a $\mathbb{Z}_n$-quantized topological invariant ~\cite{benalcazar2019quantization,BenalcazarScience,benalcazar2017quadPRB,van_Miert_2018}, where $n$ is the order of the rotation that characterizes the space group. In our case, we expect to find a $\mathbb{Z}_3$ index for the breathing phases due to the $C_{3v}$ symmetry of the lattice. In the case of the canonical kagome lattice, the bulk polarization is always zero.

In order to compute the position of the Wannier center or, equivalently, the bulk polarization, we  write the eigenvalue equation of the Wilson loop as $W_{\mathbf{b}_1}(k_2)|\nu^j_1(k_1,k_2)\rangle=\exp{\text{i}\nu^j_1(k_2)}|\nu^j_1(k_1,k_2)\rangle$, where $\nu^j_1(k_2)$ is the component of the  Wannier center of the $j$-th  Wannier function along the $\mathbf{a}_1$ direction. Taking $e=1$, and using the fact that the lattice vectors are related by the mirror $\mathbf{b}_1$, the polarization can be expressed as $\mathbf{P}=p(\mathbf{ a}_1+\mathbf{ a}_2)$, where
%
%
%%%%%%%%%%%%%%
\begin{equation}
p=\frac{1}{2\pi N_k}\sum_{j=1}^{N_k}\sum_{m=1}^{N_\text{occ}}\nu^{m}_{1,2}(k^j_{2,1}),\qquad \text{mod }1
\end{equation}
%%%%%%%%%%%%%
%
%
that is, the average of the (non-Abelian) Wilson spectrum along the reciprocal lattice vectors~\cite{lee2020fractional}. We have characterized the first three bands of the canonical and breathing kagome lattice by means of the Abelian/non-Abelian Wilson loop, respectively, since the rest of the bands are very high in energy. We obtained different values for $p$ for the three different phases. Figures~ \ref{fig:UC_choice+bands}(j) to~\ref{fig:UC_choice+bands}(n) show the  Wilson spectra obtained for the three different phases. In the case of the canonical kagome lattice, we obtain a value of $p=1/6$, which places the Wannier center at position $\mathbf{r}=(\mathbf{ a}_1+\mathbf{ a}_2)/6$. Given the basis of lattice vectors, we can state that the Wannier center is located at the 3$c$ Wyckoff position, precisely where the lattice sites are placed. Since inside the unit cell there are three equivalent 3$c$ Wyckoff positions, there are three Wannier centers located at the orbit~\footnote{The set of all the equivalent Wyckoff positions inside a unit cell is called orbit} of the 3$c$ Wyckoff position. Since the average position of the Wannier center lies at the origin of the unit cell, there is no displacement in the charge, and the polarization is thus zero (even if the computed value is above zero).

For the breathing sets of molecules, we obtained $p=0$ for the trivial phase and $p=1/3$ for the non-trivial phase. These two values allow us to locate the Wannier center at the $1b$ Wyckoff position for the trivial phase and at the $1c$ Wyckoff position for the non-trivial phase. This result is expected because in the trivial phase the intra-cell hopping is stronger than the inter-cell, and thus the surface state concentrates more around the $1b$ Wyckoff position. This results in a Wannier center placed at the origin, thus coinciding with the center of positive charge (at the $1b$ Wyckoff position). Similarly, in the non-trivial phase, the surface state concentrates more around the $1c$ Wyckoff position, yielding a negative charge center out of the center of positive charge at $1b$ Wyckoff position. In the case of the composite group, we obtain exactly the same Wannier center as the isolated band, for each of the breathing phases. The filled star in Figs.~\ref{fig:UC_choice+bands}(a),~\ref{fig:UC_choice+bands}(c), and~\ref{fig:UC_choice+bands}(e) represents the Wannier centers of the first three bands obtained via the Wilson spectrum. In Fig.~\ref{fig:UC_choice+bands}(a), the empty stars are the Wannier centers generated by the orbit of the $3c$ Wyckoff position.

The results that we have obtained are general, since we have performed the Wilson spectrum calculation using wave functions coming from the plane wave expansion of a potential, not from a tight-binding Hamiltonian. In this way, our Wilson spectra account for all possible hopping terms between lattice sites, and their behavior depends only on the symmetry properties of the lattice.

\subsection{Topological Quantum Chemistry interpretation\label{sec:4_TQC}}
%
%%%%%%%%%%%%%%
\begin{table}[!ht]
	\begin{center}
		\begin{tabular}{ccc|ccc}
			\hline
			\multicolumn{6}{|c|}{Real space}\\
			\hline
			\multicolumn{3}{c|}{$p6mm$ (\#183)}&\multicolumn{3}{c}{$p3m1$ (\#156)}\\
			MWP ($q$) & $G_q$ & Irreps&MWP ($q$) & $G_q$ & Irreps\\\hline
			\multirow{2}{*}{$3c$}&\multirow{2}{*}{$C_{2v}$}&$A_1,A_2,$&\multirow{2}{*}{$1b,\,1c$} &\multirow{2}{*}{$C_{3v}$}&\multirow{2}{*}{$A_1,A_2,E$}\\
			&&$B_1,B_2$&&&\\
			\hline\hline \\ \hline
			\multicolumn{6}{|c|}{Reciprocal space}\\
			\hline
			\multicolumn{3}{c|}{$p6mm$ (\#183)}&\multicolumn{3}{c}{$p3m1$ (\#156)}\\
			$k$ point & $G_\mathbf{k}$ & Irreps&$k$ point & $G_\mathbf{k}$ & Irreps\\ \hline
			\multirow{3}{*}{$\Gamma$}&\multirow{3}{*}{$C_{6v}$}&$A_1,A_2,$&\multirow{3}{*}{$\Gamma$} &\multirow{3}{*}{$C_{3v}$}&\multirow{3}{*}{$A_1,A_2,E$}\\
			&&$B_1,B_2$&&&\\
			&&$E_1,E_2$&&&\\
			$K,\,K'$ &$C_{3v}$&$A_1,A_2,E$&$K, K'$&$C_3$&$A,^1\!\!E,^2\!\!E$\\
			\multirow{2}{*}{$M$}&\multirow{2}{*}{$C_{2v}$}&$A_1,A_2,$&\multirow{2}{*}{$M$}&\multirow{2}{*}{$C_s$}& \multirow{2}{*}{$A',\,A''$}\\
			&&$B_1,B_2$&&&\\
			\hline\hline
		\end{tabular}
	\end{center}
	\caption{Symmetry properties of the maximal Wyckoff positions (MWP) and $k$ points involved in the quantum simulator approach of the canonical/breathing kagome lattice. Symbols $G_q$ and $G_{\bf k}$ correspond to the point groups of the Wyckoff positions and $k$ vectors, respectively.}
	\label{tab:multicol}
\end{table}

To conclude, we will use a different approach to study the topological features of a system, which is based on the symmetry eigenvalues of the Bloch wave functions at high-symmetry points in the reciprocal space. Topological Quantum Chemistry~\cite{TQC} is a powerful theory, which allows us to classify and diagnose topological phases of matter based solely on group theory arguments. Each high-symmetry point in the first Brillouin zone has a set of operations that leaves this $k$ point invariant, called little group $G_{\bf k}$. All little groups are subgroups of the full space group. For a given band structure, at each $k$ point and band index $n$, we can associate an irreducible representation (irrep) of the corresponding little group, which represents the symmetry properties of the $n$-th wave function at such $k$ point. If a set of $N$ degenerated bands touch at a certain high-symmetry point, the dimension of the associated irrep must be $N$.

The set of irreps at each $k$ point is induced by an object called band representation. A band representation is a representation of the space group that is induced ($\uparrow$) by an irrep of the point groups $G_q$ of the so-called maximal Wyckoff positions. These are the Wyckoff positions, the point group of which is a maximal subgroup of the space group. In this way, the topology of a system is fully determined by the irreps of the maximal Wyckoff positions of the space group alone. The symmetry properties are then translated from real space, by inducing ($\uparrow$) a certain band representation in real space, to reciprocal space, by particularizing the band representation at each $k$ point, a process called subduction ($\downarrow$). Elementary band representations also have dimension, and it is related as well to the number of bands conforming the whole composite group.

Once we have obtained the band representation, we may look at the position of the Wannier center from where the band representation is induced. If the Wannier center lies on an occupied maximal Wyckoff position, then the system corresponds to a trivial atomic limit, where all the hopping terms are switched off and the orbitals are unperturbed by their neighbors. On the other hand, if the Wannier center lies on an unoccupied maximal Wyckoff position, the system corresponds to an obstructed atomic limit phase.

We have characterised the canonical kagome lattice and the two breathing phases according to the symmetry eigenvalues and the Wannier centers that we obtained through the Wilson spectrum calculation. Starting from the canonical kagome lattice, the irrep assignment shown in Fig.~\ref{fig:UC_choice+bands}(g) is compatible with the three-dimensional band representation $(A_1\uparrow G)_{3c}$, induced from the $3c$ Wyckoff position. This band representation is three-dimensional because there are 3 bands touching in total. The canonical kagome lattice corresponds to a trivial atomic limit because the Wannier center lies at an occupied maximal Wyckoff positions (lattice sites). The band structure shows features that correspond to a $C_{6v}$-symmetric lattice, $i.e.$, the gap closes at $\mathbf{K}$ and $\mathbf{K}'$ points. This can be understood from symmetry arguments: the little group of the $\mathbf{K}$, $\mathbf{K}'$ points in $p6mm$ is $C_{3v}$, which shows two one-dimensional irreps ($A_1,A_2$) and a single two-dimensional irrep ($E$) (see Table \ref{tab:multicol}). For a set of three bands belonging to the $p6mm$ space group, two of them will always be degenerate due to the fact that they must transform under such two-dimensional irrep ($E$). 

After breaking the $C_6$ symmetry by introducing the breathing distortion, the symmetry of the space group is reduced to $p3m1$. In reciprocal space, the little group of the $\mathbf{K}$, $\mathbf{K}'$ points reduces from $C_{3v}$ to $C_{3}$. Since $C_{3}$ does not have two-dimensional irreps, the two-dimensional irrep from $C_{3v}$ decomposes into irreps of the new little group, which translates into a gap opening of the Dirac cones at the $\mathbf{K}$ and $\mathbf{K}'$ points. This decomposition can be studied from compatibility relations in the respective $k$ points after a symmetry reduction, revealing the pure symmetry origin of this splitting. Figure~\ref{fig:UC_choice+bands}(h) corresponds to the band structure of the trivial breathing phase with zero bulk polarization. The lowest band assignment is compatible with the band representation $(A_1\uparrow G)_{1b}$, which is one-dimensional, and the upper two are compatible with the band representation $(E\uparrow G)_{1b}$, which is two-dimensional. Both representations come from the $1b$ Wyckoff position, which is maximal, and coincides with the result obtained via the Wilson spectrum approach. Due to the fact that at this maximal Wyckoff position there is a CO molecule, this is an unnocupied maximal Wyckoff position~\footnote{The actual kagome lattice does not show any feature in the $2b/1b/1c$ maximal Wyckoff positions, neither in the canonical nor breathing phases.} and thus the phase is in an obstructed atomic limit~\cite{TQC}. Finally, Fig.~\ref{fig:UC_choice+bands}(i) shows the band structure and irrep assignment for the non-trivial phase with non-zero bulk polarization and corner states. The band representations in this case are $(A_1\uparrow G)_{1c}$ for the lowest band and $(E\uparrow G)_{1c}$ for the upper group of bands. As in the previous case, the Wannier center lies in an unoccupied maximal Wyckoff position, so the non-trivial phase corresponds to a different obstructed atomic limit.

We  find a similar setup in the SSH model: the trivial phase shows a Wannier center in the middle of the intracell link (the origin), which would correspond to a Wilson loop eigenvalue equal to zero (in 1D there  is no concept of Wilson spectrum). In contrast, the non-trivial phase shows a Wannier center on the edge of the unit cell, which corresponds to a Wilson loop eigenvalue of 1/2~\cite{bercioux2018topological}. In terms of atomic limits, the two phases are obstructed atomic limits separated by a band inversion, and thus cannot be connected adiabatically. One of them is trivial, in the sense that it displays zero bulk polarization and no corner modes, whereas the other phase is non-trivial in the sense that it displays a non-zero bulk polarization and corner modes. The following diagram shows their connection:
%
%
%%%%%%%%%%%%%
\begin{gather*}
    \text{Breathing phase with Wannier center in $1b$}\\\text{ (obstructed atomic limit, trivial bulk polarization)}\\
    \updownarrow\\
    \text{Canonical phase with Wannier centers in $3c$}\\\text{ (trivial atomic limit, trivial bulk polarization)}\\
    \updownarrow\\
    \text{Breathing phase with Wannier center in $1c$}\\\text{ (obstructed atomic limit, non-trivial bulk polarization)}
\end{gather*}
%%%%%%%%%%%%%
%
%
\section{Discussion and Conclusion}\label{sec:conclusions}
We have studied the different phases of the breathing kagome lattice from four complementary perspectives. We start by tuning a dimerization parameter that we have introduced between the intra- and inter-cell hopping terms in a tight-binding Hamiltonian. When this parameter is set to zero, we recover the traditional kagome lattice, which has a gapless spectrum. When the dimerization parameter changes sign, a band inversion occurs at the $\mathbf{K}$ and $\mathbf{K'}$ points and the two breathing phases are  distinct, while showing the same spectral properties.

A finite-size flake of the non-trivial phase has bulk, edge, and corner localized modes. These edge modes appear in the bulk gap. However, for realizing a true higher-order topological insulator, the bulk gap should host only corner modes. Hence, the breathing kagome lattice does not encode higher-order topology. To study the origin, symmetries, and properties protecting such corner modes, we have introduced several perturbations to the non-trivial phase of a finite-size triangular-shaped flake using a tight-binding formalism. We have chosen this geometry to ensure that the sample respects the $C_{3v}$ symmetry group of the lattice. We have shown that the corner modes are trivial, and that three ingredients are needed to pin the modes to zero-energy, and to localize a corner state at one sublattice. First of all, the symmetries imposed by the space group should be respected. Breaking spatial symmetries would lead to, for example, non-degenerate corner modes, as we saw by introducing random on-site energies in the flake, while respecting the kagome pattern (see Fig.~\ref{figure_two}(a)). Second, we cannot connect sites belonging to the same sublattice, i.e., this is the same as preserving generalized chiral symmetry, which strongly affects the way in which the modes move away from zero, as we saw in Fig.~\ref{figure_two}(b). Finally, the connectivity between lattice sites of different species must be done in a consecutive way, constructing a closed triangle of vertices ABC (see Fig.~\ref{figure_two}(e)).  Importantly, if and only if these conditions are fulfilled, the corner modes are truly localized in the corner sublattice, tightly pinned to zero. We have also confirmed that these rules can be extended up to second, third, etc., nearest neighbors while increasing accordingly the size of the flakes. Otherwise, the corner modes would move away from zero due to overlap.

We have found similar rules of localization of corner modes after studying a system which resembles the breathing kagome lattice: a kagome rhombus that displays the same sublattice in each corner. A destructive interference solution can be found if the corner sublattice is connected to the rest of the system according to certain rules, which are equivalent to the rules stated in the previous paragraph, i.e, both approaches give the same result.

Finally, we have performed a study of the kagome lattice based on a muffin-tin calculation. In this picture, with no concept of individual hopping terms, all possible overlap between all the lattice sites are included in the calculation. By solving the Schr\"odinger equation, we obtained the Bloch wave functions, which inherit all the symmetry properties from the periodic potential. After applying a Wilson spectrum characterization and symmetry markers, we have been able to identify the band representation to which each phase corresponds (canonical, trivial, and non-trivial). We found that the two breathing phases correspond to two different obstructed atomic limits, connected through a gap closing. Hence, these two phases are not adiabatically connected. This gap closing reveals a band inversion between the two phases. It also accounts for the recovery of a six-fold rotation, characteristic of the canonical kagome lattice. This setup actually corresponds to a  trivial atomic limit in which, up to a point group operation, the Wannier centers lie exactly at the lattice sites. 

These results may shed light on the protection of the corner modes of two-dimensional lattices, as well as on  understanding of what is  and what it is not a HOTI. Within a more general framework than a tight-binding Hamiltonian, we have demonstrated the trivial/non-trivial distinction between the two phases of the breathing kagome lattice, as well as the source of the existence and protection of the corner modes. Since the muffin-tin technique accounts for all the possible hopping terms between all lattice sites, we believe that both the existence and protection of corner modes are a consequence of the symmetry properties of the non-trivial phase hosting the corner modes. In addition, the Wilson spectrum characterization of all the phases of the kagome lattice is determined exclusively by the symmetries of the lattice. However, the appearance of edge modes in the bulk gap of the finite-size system suggests that this protection does not have any topological character, while being robust to some extent. We conclude that the corner modes of the breathing kagome lattice have some robustness but are not topological.

Robust protection of corner modes may have potential applications for lasing techniques~\cite{kim2020lasing,kagomewaveguide,El_Hassan_2019}. These references are based on a kagome pattern, so we believe that the corner modes that they propose do not possess any topological protection, while being robust by the symmetry of the lattice.

\vspace{1cm}

\section{Acknowledgements\label{sec:acknowledgements}}
We acknowledge useful discussions with Wouter Beugeling, Barry Bradlyn, Maia Garcia Vergniory, Flore Kunst, Mikel Iraola, Titus Neupert, Jette van den Broeke and Robin Verstraten.  The work of M.A.J.H. and D.B. is supported by the Ministerio de Ciencia e Innovaci\'on  (MICINN) through Project No.~PID2020-120614GB-I00, and by the Transnational Common Laboratory $Quantum-ChemPhys$ (D.B.). 
A.G.E. and M.B.P. acknowledge support from the Spanish Ministerio de Ciencia e Innovaci\'on (Project No.~PID2019-109905GA-C2) and from Eusko Jaurlaritza (Grants No.~IT1164-19 and No.~KK-2021/00082). A.G.E. and D.B acknowledge Programa Red Guipuzcoana de Ciencia, Tecnolog\'ia e Innovaci\'on 2021, Grant No.~2021-CIEN-000070-01 Gipuzkoa Next. A.G.E., M.B.P. and D. B. acknowledge funding from the Basque Government's IKUR initiative on Quantum technologies (Department of Education).
I.S. gratefully acknowledges financial support from the European Research Council (Horizon 2020  ``FRACTAL", Grant No.~865570).

\appendix

\section{Tight-binding model of the breathing kagome lattice}\label{appA}
Here, we show an explicit calculation for a similar kagome lattice as shown in Fig.~\ref{figure_kagome_rhombus} in the main text, consisting of 21 sites to keep the equation concise.

A solution for the equation $H \psi = e_\text{A} \psi$ is found if $t_\text{BA}^{m'} = t_\text{BA}^{mm'}= t_\text{BA}^{mm''} = t_\text{AB}^{m'} = t_\text{CA}^m =  t_\text{AC}^m = t_\text{AA}^m=t_\text{AA}^{m'}= t_\text{AA}^{mm'}= t_\text{AB}^{mm'} =t_\text{AB}^{mm''} = t_\text{AC}^{mm'} =t_\text{AC}^{mm''} =  t_\text{CA}^{mm'}= t_\text{CA}^{mm''}=0$. These hopping values are indicated by in green in Fig.~\ref{figure_kagome_rhombus} of the main text. In the following, we decompose the $21\times21$ Hamiltonian matrix $H$ into a set of 9 $M$ matrices of dimension $7\times7$:
%
%
%%%%%%%%%%%%%
\begin{align}
    \begin{pmatrix}
      M_{11} & M_{12} & M_{13} \\
      M_{21} & M_{22} & M_{32} \\
      M_{13} & M_{32} & M_{33} 
    \end{pmatrix}
      \psi =
   \chi,
\end{align}
%%%%%%%%%%%%%
%
%
where each matrix is defined as:
\begin{widetext}
%
%
%%%%%%%%%%%
\begin{subequations}

\begin{align}
    & M_{11}= \begin{pmatrix}
       e_\text{A} & -t_\text{AB} & -t_\text{AA}^m & -t_\text{AB}^m & 0 & -t_\text{AC} & -t_\text{AC}^m \\
       -t_\text{AB} & e_\text{B} & -t_\text{BA} & -t_\text{BB}^m & -t_\text{BA}^m & -t_\text{BC} & -t_\text{BC}^m \\
       -t_\text{AA}^m & -t_\text{BA} & e_\text{A} & -t_\text{AB} & -t_\text{AA}^m & -t_\text{CA}^m & -t_\text{AC} \\
       -t_\text{AB}^m & -t_\text{BB}^m & -t_\text{AB} & e_\text{B} & -t_\text{BA} & -t_\text{CB}^m & -t_\text{BC} \\
       0 & -t_\text{BA}^m & -t_\text{AA}^m & -t_\text{BA} & e_\text{A} & 0 & -t_\text{CA}^m \\
       -t_\text{AC} & -t_\text{BC} & -t_\text{CA}^m & -t_\text{CB}^m & 0 & e_C & -t_\text{CC} \\
       -t_\text{AC}^m & -t_\text{BC}^m & -t_\text{AC} & -t_\text{BC} & -t_\text{CA}^m & -t_\text{CC} & e_\text{C}
    \end{pmatrix},
    \end{align}
    \begin{align}
    & M_{12} = \begin{pmatrix}
      0 & -t_\text{AA}^{m'} & -t_\text{AB}^{m'} & 0 & 0 & 0 & -t_\text{AC}^{m'}\\
      0 & -t_\text{BA}^{m'} & -t_\text{BB}^{m'} & -t_\text{BA}^{mm''} & 0 & 0 & -t_\text{BC}^{m'} \\
      -t_\text{AC}^m & -t_\text{AA}^{mm'} & -t_\text{AB}^{mm'} & -t_\text{AA}^{m'} & -t_\text{AB}^{m'} & 0 & -t_\text{AC}^{mm'} \\
      -t_\text{BC}^m & -t_\text{BA}^{mm'} & -t_\text{BB}^{mm'} & -t_\text{BA}^{m'} & -t_\text{BB}^{m'} & -t_\text{BA}^{mm''} & -t_\text{BC}^{mm'} \\
      -t_\text{AC} & 0 & -t_\text{AB}^{mm''} & -t_\text{AA}^{mm'} & -t_\text{AB}^{mm'} & -t_\text{AA}^{m'} & 0 \\
      0 & -t_\text{CA} & -t_\text{CB}^{m'} & -t_\text{CA}^{mm''} & 0 & 0 & -t_\text{CC}^{m'} \\
      -t_\text{CC} & -t_\text{CA}^{m m'} & -t_\text{CB} & -t_\text{CA} & -t_\text{CB}^{mm'} & -t_\text{CA}^{mm''} & -t_\text{CC}^{mm'}
    \end{pmatrix},
        \end{align}
    \begin{align}
    & M_{13} = \begin{pmatrix}
      0 & 0 & 0 & 0 & 0 & 0 & 0 \\
      0 & 0 & 0 & 0 & 0 & 0 & 0 \\
      -t_\text{AC}^{m'} & 0 & 0 & 0 & 0 & 0 & 0 \\
      -t_\text{BC}^{m'} & 0 & 0 & 0 & 0 & 0 & 0\\
      -t_\text{AC}^{mm'} & -t_\text{AC}^{m'} & 0 & 0 & 0 & 0 & 0 \\
      0 & 0 & -t_\text{CA}^{m'} & 0 & 0 & 0 & 0\\
      -t_\text{CC}^{m'} & 0 & -t_\text{AC}^{mm''} & -t_\text{CB}^{mm^{(3)}} & -t_\text{CA}^{m'} & 0 & 0
    \end{pmatrix},
\end{align}
\begin{align}
    & M_{21}=\begin{pmatrix}
      0 & 0 & -t_\text{AC}^m & -t_\text{BC}^m & -t_\text{AC} & 0 & -t_\text{CC}\\
      -t_\text{AA}^{m'} & -t_\text{BA}^{m'} & -t_\text{AA}^{mm'} & -t_\text{BA}^{mm'} & 0 & -t_\text{CA} & -t_\text{CA}^{m m'} \\
      -t_\text{AB}^{m'} & -t_\text{BB}^{m'} & -t_\text{AB}^{mm'} & -t_\text{BB}^{mm'} & -t_\text{AB}^{mm''} & -t_\text{CB}^{m'} & -t_\text{CB} \\
      0 & -t_\text{BA}^{mm''} & -t_\text{AA}^{m'} & -t_\text{BA}^{m'} & -t_\text{AA}^{mm'} & -t_\text{CA}^{mm''} & -t_\text{CA}\\
       0 & 0 & -t_\text{AB}^{m'} & -t_\text{BB}^{m'} & -t_\text{AB}^{mm'} & 0 & -t_\text{CB}^{mm'}\\
       0 & 0 & 0 & -t_\text{BA}^{mm''} & -t_\text{AA}^{m'} & 0 & -t_\text{CA}^{mm''} \\
       -t_\text{AC}^{m'} & -t_\text{BC}^{m'} & -t_\text{AC}^{mm'} & -t_\text{BC}^{mm'} & 0 & -t_\text{CC}^{m'} & -t_\text{CC}^{mm'}
    \end{pmatrix},
    \end{align}
    \begin{align}
    & M_{22}=\begin{pmatrix}
      e_\text{C} & 0 & -t_\text{CB}^{mm''} & -t_\text{CA}^{m m'} & -t_\text{CB} & -t_\text{CA} & 0 \\
      0 & e_\text{A} & -t_\text{AB} & -t_\text{AA}^m & -t_\text{AB}^m & 0 & -t_\text{AC} \\
      -t_\text{CB}^{mm''} & -t_\text{AB} & e_\text{B} & -t_\text{BA} & -t_\text{BB}^m & -t_\text{BA}^m & -t_\text{BC}\\
      -t_\text{CA}^{m m'} & -t_\text{AA}^m & -t_\text{BA} & e_\text{A} & -t_\text{AB} & -t_\text{AA}^m & -t_\text{CA}^m  \\
      -t_\text{CB} & -t_\text{AB}^m & -t_\text{BB}^m & -t_\text{AB} & e_\text{B} & -t_\text{BA} & -t_\text{CB}^m\\% this should not be here& -t_\text{BC} \\
      -t_\text{CA} & 0 & -t_\text{BA}^m & -t_\text{AA}^m & -t_\text{BA} & e_\text{A} & 0 \\
      0 & -t_\text{AC} & -t_\text{BC} & -t_\text{CA}^m & -t_\text{CB}^m & 0 & e_\text{C} \\
    \end{pmatrix},
    \end{align}
    \begin{align}
    & M_{23}=\begin{pmatrix}
      -t_\text{CC}^{mm'} & -t_\text{CC}^{m'} & 0 & -t_\text{BC}^{mm''} & -t_\text{AC}^{mm''} & -t_\text{CB}^{mm^{(3)}} & -t_\text{CA}^{m'} \\
      -t_\text{AC}^m & 0 & -t_\text{AA}^{m'} & -t_\text{AB}^{m'} & 0 & 0 & 0 \\
      -t_\text{BC}^m & 0 & -t_\text{BA}^{m'} & -t_\text{BB}^{m'} & -t_\text{BA}^{mm''} & 0 & 0 \\
      -t_\text{AC} & -t_\text{AC}^m & -t_\text{AA}^{mm'} & -t_\text{AB}^{mm'} & -t_\text{AA}^{m'} & -t_\text{AB}^{m'} & 0 \\
      -t_\text{BC}^m & -t_\text{BA}^{mm'} & -t_\text{BB}^{mm'} & -t_\text{BA}^{m'} & -t_\text{BB}^{m'} & -t_\text{BA}^{mm''} & 0\\
      -t_\text{CA}^m & -t_\text{AC} & 0 & -t_\text{AB}^{mm''} & -t_\text{AA}^{mm'} & -t_\text{AB}^{mm'} & -t_\text{AA}^{m'} \\
      -t_\text{CC} & 0 & -t_\text{CA} & -t_\text{CB}^{m'} & -t_\text{CA}^{mm''} & 0 & 0
    \end{pmatrix},
\end{align}
and
\begin{align}
        & M_{31}=\begin{pmatrix}
        0 & 0 & -t_\text{AC}^{m'} & -t_\text{BC}^{m'} & -t_\text{AC}^{mm'} & 0 & -t_\text{CC}^{m'} \\
        0 & 0 & 0 & 0 & -t_\text{AC}^{m'} & 0 & 0 \\
        0 & 0 & 0 & 0 & 0 & -t_\text{CA}^{m'} & -t_\text{AC}^{mm''} \\
        0 & 0 & 0 & 0 & 0 & 0 & -t_\text{CB}^{mm^{(3)}} \\
        0 & 0 & 0 & 0 & 0 & 0 & -t_\text{CA}^{m'} \\% this should not be here&& -t_\text{AC}^{mm''} \\
        0 & 0 & 0 & 0 & 0 & 0 & 0 \\
        0 & 0 & 0 & 0 & 0 & 0 & 0
        \end{pmatrix},
        \end{align}
        \begin{align}
        & M_{32}=\begin{pmatrix}
        -t_\text{CC}^{mm'} & -t_\text{AC}^m & -t_\text{BC}^m & -t_\text{AC} & -t_\text{BC} & -t_\text{CA}^m & -t_\text{CC} \\
        -t_\text{CC}^{m'} & 0 & 0 & -t_\text{AC}^m & -t_\text{BC}^m & -t_\text{AC} & 0 \\
        0 & -t_\text{AA}^{m'} & -t_\text{BA}^{m'} & -t_\text{AA}^{mm'} & -t_\text{BA}^{mm'} & 0 & -t_\text{CA} \\
        -t_\text{BC}^{mm''} & -t_\text{AB}^{m'} & -t_\text{BB}^{m'} & -t_\text{AB}^{mm'} & -t_\text{BB}^{mm'} & -t_\text{AB}^{mm''} & -t_\text{CB}^{m'} \\
        0 & -t_\text{BA}^{mm''} & -t_\text{AA}^{m'} & -t_\text{BA}^{m'} & -t_\text{AA}^{mm'} & -t_\text{CA}^{mm''}&0\\
        -t_\text{CB}^{mm^{(3)}} & 0 & 0 & -t_\text{AB}^{m'} & -t_\text{BB}^{m'} & -t_\text{AB}^{mm'} & 0 \\
        -t_\text{CA}^{m'} & 0 & 0 & 0 & -t_\text{BA}^{mm''} & -t_\text{AA}^{m'} & 0
        \end{pmatrix},
        \end{align}
        \begin{align}
        & M_{33}=\begin{pmatrix}
        e_\text{C} & -t_\text{CC} & -t_\text{CA}^{m m'} & -t_\text{CB} & -t_\text{CA} & -t_\text{CB}^{mm'} & -t_\text{CA}^{mm''} \\
        -t_\text{CC} & e_\text{C} & 0 & -t_\text{CB}^{mm''} & -t_\text{CA}^{m m'} & -t_\text{CB} & -t_\text{CA} \\
        -t_\text{CA}^{m m'} & 0 & e_\text{A} & -t_\text{AB} & -t_\text{AA}^m & -t_\text{AB}^m & 0 \\
        -t_\text{CB} & -t_\text{CB}^{mm''} & -t_\text{AB} & e_\text{B} & -t_\text{BA} & -t_\text{BB}^m & -t_\text{BA}^m \\
        -t_\text{CA} & -t_\text{CA}^{m m'} & -t_\text{AA}^m & -t_\text{BA} & e_\text{A} & -t_\text{AB} & -t_\text{AA}^m \\
        -t_\text{CB}^{mm'} & -t_\text{CB} & -t_\text{AB}^m & -t_\text{BB}^m & -t_\text{AB} & e_\text{B} & -t_\text{BA}\\
        -t_\text{CA}^{mm''} & -t_\text{CA} & 0 & -t_\text{BA}^m & -t_\text{AA}^m & -t_\text{BA} & e_\text{A}
        \end{pmatrix}.
\end{align}
\end{subequations}
%%%%%%%%%%%
%
%
The vector ansatz for the localized state $\psi$ reads
%
%
%%%%%%%%%%
\begin{equation}
    \psi = \left( 1 \quad 0 \quad r_1 \quad 0 \quad r_2\quad 0\quad  0 \quad 0  \quad r_3 \quad 0\quad  r_4 \quad  0 \quad r_5 \quad 0 \quad 0 \quad 0 \quad r_6 \quad 0 \quad r_7 \quad 
 0 \quad r_8
    \right)^\text{T},
\end{equation}
and finally, the action of the system Hamiltonian of the ansatz vector is given by $\chi=(\chi_1 \quad \chi_2\quad \chi_3)^\text{T}$:
%
%
%%%%%%%%%%%%
\begin{subequations}
\begin{align}
    \chi_1 = \begin{pmatrix}
       e_\text{A}-r_3 t_\text{AA}^{m'}-r_1 t_\text{AA}^m \\
 -t_\text{AB}-r_3 t_\text{BA}^{m'}-r_2 t_\text{BA}^m-r_4 t_\text{BA}^{mm''}-r_1 t_\text{BA} \\
 r_1 e_\text{A}-r_4 t_\text{AA}^{m'}-r_2 t_\text{AA}^m-t_\text{AA}^m-r_3 t_\text{AA}^{mm'} \\
 -t_\text{AB}^m-r_1 t_\text{AB}-r_4 t_\text{BA}^{m'}-r_5 t_\text{BA}^{mm''}-r_3 t_\text{BA}^{mm'}-r_2 t_\text{BA} \\
 r_2 e_\text{A}-r_5 t_\text{AA}^{m'}-r_1 t_\text{AA}^m-r_4 t_\text{AA}^{mm'} \\
 -t_\text{AC}-r_6 t_\text{CA}^{m'}-r_1 t_\text{CA}^m-r_4 t_\text{CA}^{mm''}-r_3 t_\text{CA} \\
 -t_\text{AC}^m-r_6 t_\text{AC}^{mm''}-r_1 t_\text{AC}-r_7 t_\text{CA}^{m'}-r_3 t_\text{CA}^{m m'}-r_2 t_\text{CA}^m-r_5 t_\text{CA}^{mm''}-r_4 t_\text{CA} 
    \end{pmatrix},
    \end{align}
    \begin{align}
    \chi_2 = \begin{pmatrix}
       -r_1 t_\text{AC}^m-r_7 t_\text{AC}^{mm''}-r_2 t_\text{AC}-r_8 t_\text{CA}^{m'}-r_4 t_\text{CA}^{m m'}-r_5 t_\text{CA} \\
 r_3 e_\text{A}-r_6 t_\text{AA}^{m'}-t_\text{AA}^{m'}-r_4 t_\text{AA}^m-r_1 t_\text{AA}^{mm'} \\
 -t_\text{AB}^{m'}-r_2 t_\text{AB}^{mm''}-r_1 t_\text{AB}^{mm'}-r_3 t_\text{AB}-r_6 t_\text{BA}^{m'}-r_5 t_\text{BA}^m-r_7 t_\text{BA}^{mm''}-r_4 t_\text{BA} \\
 r_4 e_\text{A}-r_1 t_\text{AA}^{m'}-r_7 t_\text{AA}^{m'}-r_3 t_\text{AA}^m-r_5 t_\text{AA}^m-r_2 t_\text{AA}^{mm'}-r_6 t_\text{AA}^{mm'} \\
 -r_1 t_\text{AB}^{m'}-r_3 t_\text{AB}^m-r_2 t_\text{AB}^{mm'}-r_4 t_\text{AB}-r_7 t_\text{BA}^{m'}-r_8 t_\text{BA}^{mm''}-r_6 t_\text{BA}^{mm'}-r_5 t_\text{BA} \\
 r_5 e_\text{A}-r_2 t_\text{AA}^{m'}-r_8 t_\text{AA}^{m'}-r_4 t_\text{AA}^m-r_7 t_\text{AA}^{mm'} \\
 -t_\text{AC}^{m'}-r_1 t_\text{AC}^{mm'}-r_3 t_\text{AC}-r_4 t_\text{CA}^m-r_7 t_\text{CA}^{mm''}-r_6 t_\text{CA}
    \end{pmatrix},
    \end{align}
    \begin{align}
    \chi_3 = \begin{pmatrix}
       -r_1 t_\text{AC}^{m'}-r_3 t_\text{AC}^m-r_2 t_\text{AC}^{mm'}-r_4 t_\text{AC}-r_6 t_\text{CA}^{m m'}-r_5 t_\text{CA}^m-r_8 t_\text{CA}^{mm''}-r_7 t_\text{CA} \\
 -r_2 t_\text{AC}^{m'}-r_4 t_\text{AC}^m-r_5 t_\text{AC}-r_7 t_\text{CA}^{m m'}-r_8 t_\text{CA} \\
 r_6 e_\text{A}-r_3 t_\text{AA}^{m'}-r_7 t_\text{AA}^m-r_4 t_\text{AA}^{mm'} \\
 -r_3 t_\text{AB}^{m'}-r_5 t_\text{AB}^{mm''}-r_4 t_\text{AB}^{mm'}-r_6 t_\text{AB}-r_8 t_\text{BA}^m-r_7 t_\text{BA} \\
 r_7 e_\text{A}-r_4 t_\text{AA}^{m'}-r_6 t_\text{AA}^m-r_8 t_\text{AA}^m-r_5 t_\text{AA}^{mm'} \\
 -r_4 t_\text{AB}^{m'}-r_6 t_\text{AB}^m-r_5 t_\text{AB}^{mm'}-r_7 t_\text{AB}-r_8 t_\text{BA} \\
 r_8 e_\text{A}-r_5 t_\text{AA}^{m'}-r_7 t_\text{AA}^m
    \end{pmatrix}. 
\end{align}
\end{subequations}
%%%%%%%%%%
%
%
\end{widetext}

%
%
%%%%%%%%%%%%%%%%%%%%
\begin{figure*}[!ht]
    \centering\includegraphics[width=0.75\linewidth]{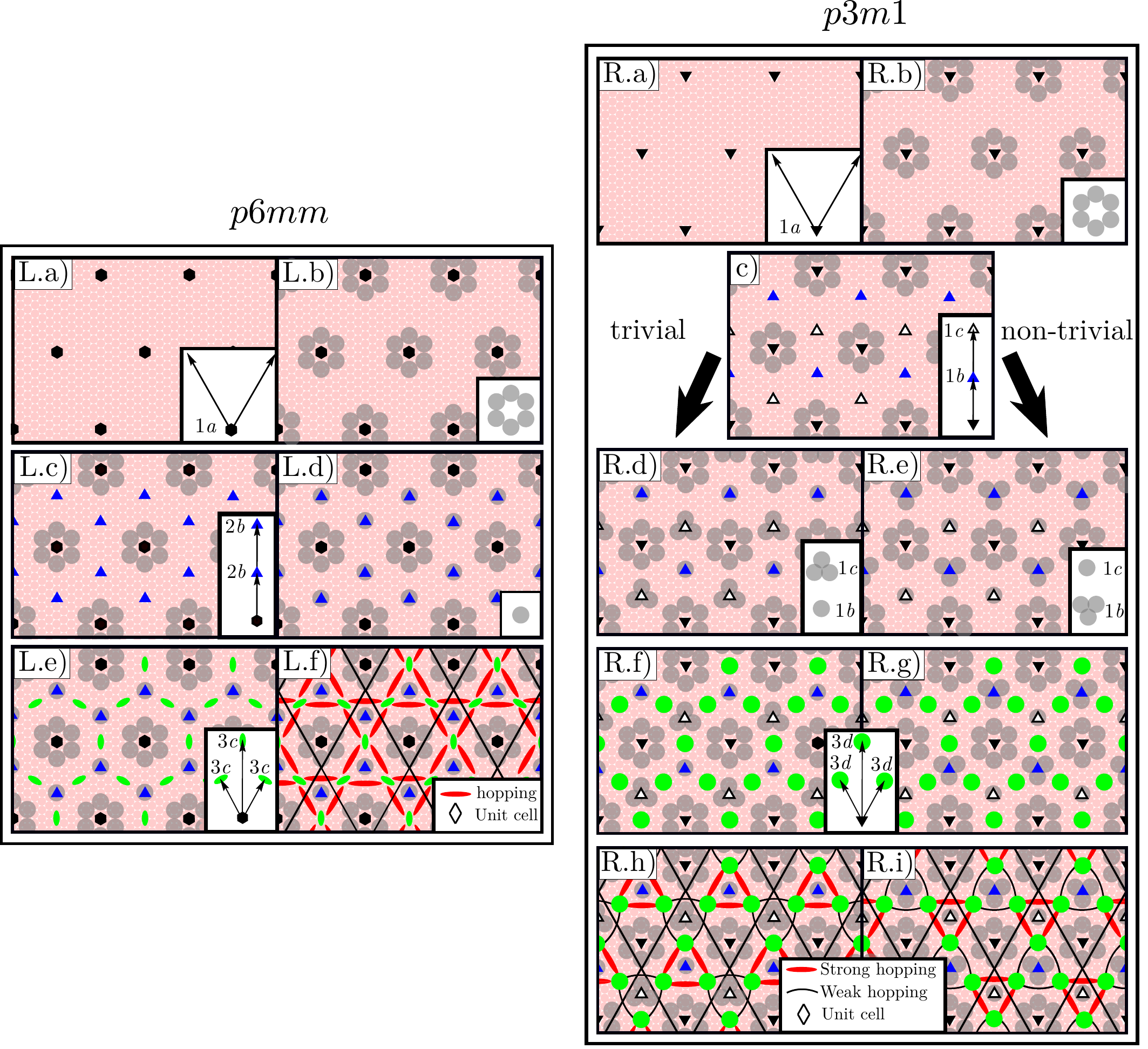}
    \caption{Steps for realizing  the muffin-tin potential for the kagome lattice in the 
    canonical phase (left panel) and the two breathing phases (right panel)  using CO molecules.}
    \label{fig_app_a}
\end{figure*}
%%%%%%%%%%%%%%
%
%
\section{Steps for building muffin-tin potentials for the canonical/breathing kagome lattice\label{App:instructions}}
In this appendix we present a step-by-step procedure for realizing the kagome lattice within the muffin-tin technique, both for the canonical and the breathing phases, using CO molecules on top of the Cu (111) surface.

\subsection{Canonical case}
We begin by studying the geometry of the kagome lattice in terms of Wyckoff positions. In the canonical form, the kagome lattice is a triangular lattice belonging to the $p6mm$ plane space group. Such space group has the following maximal Wyckoff positions: $1a$, $2b$, $3c$. In the case of the kagome lattice, the lattice sites are the $3c$ Wyckoff position, and the remaining one are unoccupied. To realize the muffin-tin potential, we place CO molecules to block the wave function from localizing in the unoccupied Wyckoff positions. Thus, we have placed six CO molecules forming an hexagon around the $1a$ Wyckoff position and a single molecule on the $2b$ Wyckoff position, thus leaving free the $3c$ Wyckoff position. In this way, the 2DEG will be confined just on the lattice formed by the $3c$ Wyckoff positions, thus reproducing the canonical kagome lattice. The left panel of Fig.~\ref{fig_app_a} represents this process step by step. To be consistent with the text, we have represented Wyckoff positions with elements with the same point group symmetry as the Wyckoff position.
%
%
%%%%%%%%%%%%%%%%%
\begin{figure*}[!th]
    \centering
    \includegraphics[width=.7\textwidth]{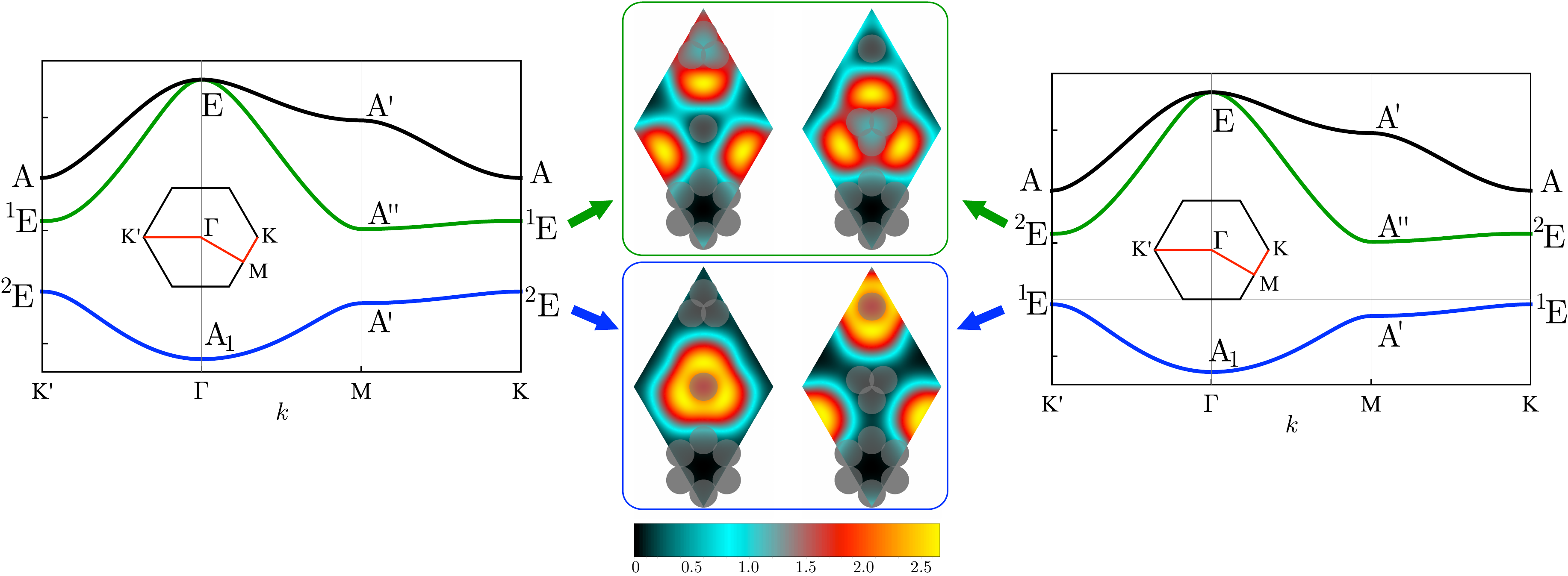}
    \caption{Wave functions for the first and second bands of the trivial (left) and non-trivial (right) configurations of the muffin-tin setup.}
    \label{fig:wavefunc}
\end{figure*}
%%%%%%%%%%%%%%
%
%
\subsection{Breathing distortion}
When we introduce the breathing distortion in the kagome lattice, we break the $C_6$ symmetry, so the space group is reduced from $p6mm$ to $p3m1$, one of its subgroups. This group/subgroup relation splits the $2b$ Wyckoff position into $1b$ and $1c$, which are non-equivalent. Additionally, the symmetry of the $3c$ Wyckoff position, now called $3d$, reduces from $C_{2v}$ to $C_m$. The feature showing in each of them is what defines the trivial/non-trivial phase. The $3d$ Wyckoff positions enclose the $1b$ Wyckoff position, so if we place a single molecule in the $1b$, and three molecules in the $1c$, we make the effective intercell hopping amplitude smaller than the intracell one. This situation corresponds to the trivial case, which does not show corner modes. The non-trivial case can be achieved by inverting the $1b$ and $1c$ Wyckoff positions. Now the $3d$ Wyckoff positions have a smaller effective intracell hopping amplitude compared to the intercell one, so the corners would host zero energy modes since they are weakly connected to the rest of the lattice. Again, we have confined the 2DEG to a lattice formed by the $3d$ Wyckoff positions, thus reproducing the breathing kagome lattice. The right panel of Fig.~\ref{fig_app_a} represents this process step by step. To be consistent with the text, we have represented Wyckoff positions with elements with the same point group symmetry as the Wyckoff position.

\section{Wave functions for trivial/non-trivial setups}

Once we have solved the Schr\"odinger equation, we can reconstruct the Bloch wave function and plot it in real space. We show in Fig.~\ref{fig:wavefunc} the modulo squared of the wave functions for the first two bands at $\mathbf{K},\,\mathbf{K'}$ points, where the band inversion occurs. The left and right panels show the band structures and irrep assignments of the trivial and non-trivial phases of the breathing kagome lattice in the muffin-tin setup. The middle panel show the plot of the wave function inside the unit cell for the first two bands right at the point where the band inversion occurs. We can see that the wave functions transforming as the $^2E$ irrep looks like a triangle pointing up, while the wave functions transforming as the $^1E$ irrep resembles a triangle pointing down plus a translation of $(\mathbf{a}_1+\mathbf{a}_2)/3$.

\newpage
\bibliographystyle{apsrev4-1} 
\bibliography{bib}  

\end{document}